\documentclass[final,5p,times,twocolumn,authoryear]{elsarticle}

\usepackage{epsfig}

\usepackage{natbib}

\usepackage{amssymb}

\usepackage{lineno}
\usepackage{rotating}
\usepackage[bookmarks = true, bookmarksnumbered = true, pdfpagemode =None, pdfstartview = FitH, pdfpagelayout = SinglePage, colorlinks = true, urlcolor = blue, citecolor = monbleu]{hyperref}

\newcommand{\degree}{\ensuremath{^\circ}}

\journal{Icarus}

\begin{document}

\begin{frontmatter}


\title{Jovian Temperature and Cloud Variability during the 2009-2010 Fade of the South Equatorial Belt}

\author[ox,le]{Leigh N. Fletcher}
\ead{leigh.fletcher@leicester.ac.uk}
\author[jpl]{G.S. Orton}
\author[jhr]{J.H. Rogers}
\author[gsfc]{A. A. Simon-Miller}
\author[berk]{I. de Pater}
\author[berk]{M.H. Wong}
\author[bes]{O. Mousis}
\author[ox]{P.G.J. Irwin}
\author[mj]{M. Jacquesson}
\author[jpl]{P.A. Yanamandra-Fisher}




\address[ox]{Atmospheric, Oceanic \& Planetary Physics, Department of Physics, University of Oxford, Clarendon Laboratory, Parks Road, Oxford, OX1 3PU, UK}
\address[le]{Department of Physics \& Astronomy, University of Leicester, University Road, Leicester, LE1 7RH, UK}
\address[jpl]{Jet Propulsion Laboratory, California Institute of Technology, 4800 Oak Grove Drive, Pasadena, CA, 91109, USA}
\address[jhr]{British Astronomical Association, Burlington House, Piccadilly, London W1J ODU, UK}
\address[gsfc]{NASA/Goddard Spaceflight Center, Greenbelt, Maryland, 20771, USA }
\address[berk]{University of California, Berkeley, Astronomy Dept., 601 Campbell Hall, Berkeley, CA 94720-3411, USA}
\address[bes]{Institut UTINAM, CNRS-UMR 6213, Observatoire de Besan\c{c}on,  Universit{\'e} de Franche-Comt{\'e}, Besan\c{c}on, France}
\address[mj]{JUPOS Team, C/O British Astronomical Association, Burlington House, Piccadilly, London W1J ODU, UK.}


\begin{abstract}

Mid-infrared 7-20 $\mu$m imaging of Jupiter from ESO's Very Large Telescope (VLT/VISIR) demonstrate that the increased albedo of Jupiter's South Equatorial Belt (SEB) during the `fade' (whitening) event of 2009-2010 was correlated with changes to atmospheric temperature and aerosol opacity.    The opacity of the tropospheric condensation cloud deck at pressures less than 800 mbar increased by 80\% between May 2008 and July 2010, making the SEB (7-17\degree S) as opaque in the thermal infrared as the adjacent equatorial zone.   After the cessation of discrete convective activity within the SEB in May 2009, a cool band of high aerosol opacity (the SEB zone at 11-15\degree S) was observed separating the cloud-free northern and southern SEB components.  The cooling of the SEBZ (with peak-to-peak contrasts of $1.0\pm0.5$ K), as well as the increased aerosol opacity at 4.8 and 8.6 $\mu$m, preceded the visible whitening of the belt by several months.  A chain of five warm, cloud-free `brown barges' (subsiding airmasses) were observed regularly in the SEB between June 2009 and June 2010, by which time they too had been obscured by the enhanced aerosol opacity of the SEB, although the underlying warm circulation was still present in July 2010.  Upper tropospheric temperatures (150-300 mbar) remained largely unchanged during the fade, but the cool SEBZ formation was detected at deeper levels ($p>300$ mbar) within the convectively unstable region of the troposphere.  The SEBZ formation caused the meridional temperature gradient of the SEB to decrease between 2008 and 2010, reducing the vertical thermal windshear on the zonal jets bounding the SEB.  The southern  SEB had fully faded by July 2010 and was characterised by short-wave undulations at 19-20\degree S.  The northern SEB persisted as a narrow grey lane of cloud-free conditions throughout the fade process.  The cool temperatures and enhanced aerosol opacity of the SEBZ after July 2009 are consistent with an upward flux of volatiles (e.g., ammonia-laden air) and enhanced condensation, obscuring the blue-absorbing chromophore and whitening the SEB by April 2010.  These changes occurred within cloud decks in the convective troposphere, and not in the radiatively-controlled upper troposphere.  NH$_3$ ice coatings on aerosols at $p<800$ mbar are plausible sources of the suppressed 4.8 and 8.6-$\mu$m emission, although differences in the spatial distribution of opacity at these two wavelengths suggest that enhanced attenuation by a deeper cloud ($p>800$ mbar) also occurred during the fade.  Revival of the dark SEB coloration in the coming months will ultimately require sublimation of these ices by subsidence and warming of volatile-depleted air.

\end{abstract}


\end{frontmatter}


\section{Introduction}
\label{intro}

\begin{figure*}[tbp]
\centering
\epsfig{file=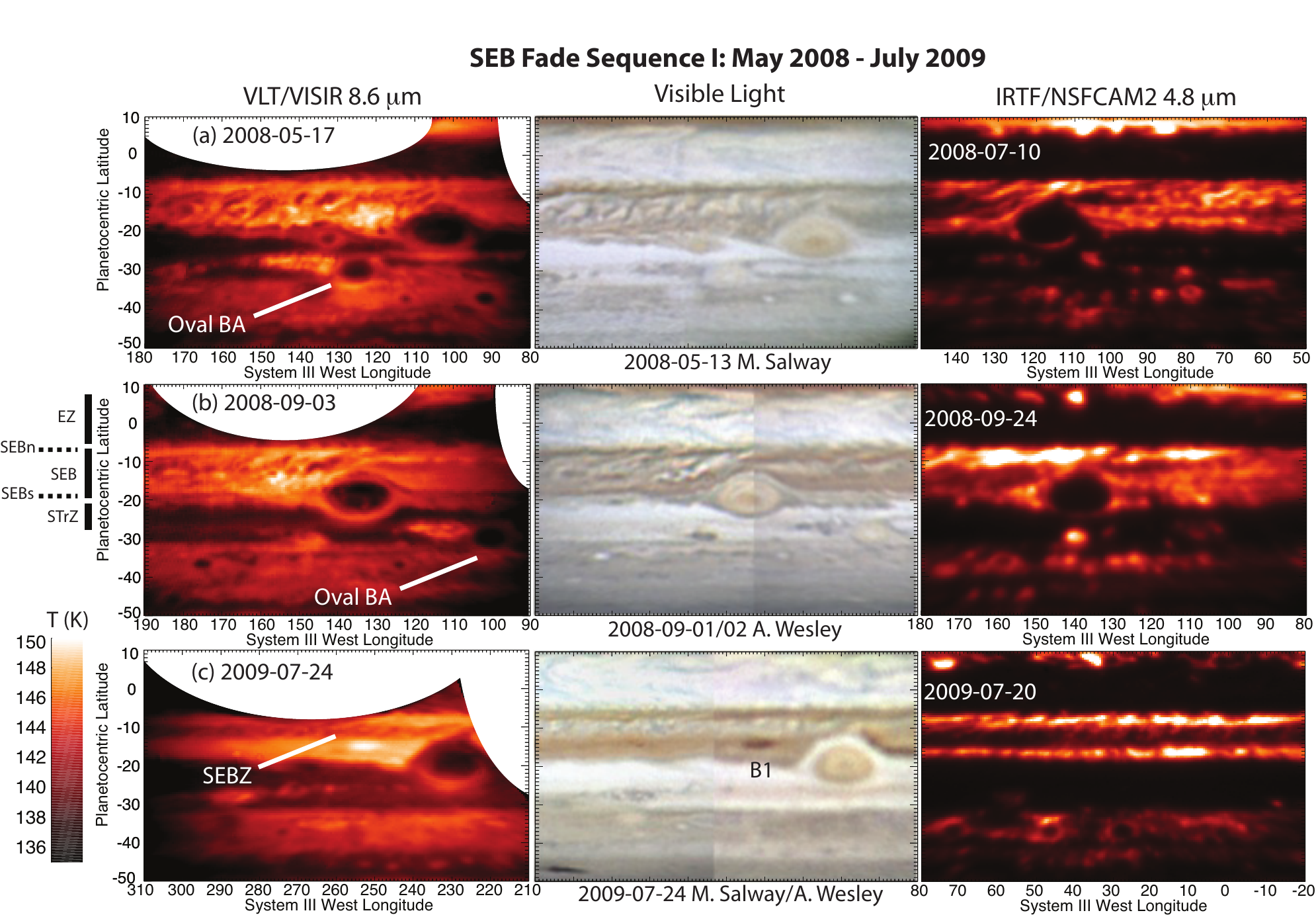,width=\textwidth,angle=0}
\caption{The SEB fade sequence May 2008-July 2009, observed by VLT/VISIR at 8.6 $\mu$m (left, sensitive to aerosol opacity at $p<800$ mbar); visible light from amateur observers (centre); and IRTF/NSFCAM2 observations at 4.8 $\mu$m (right, sensitive to aerosols above the 2-3 bar level).  Visible images taken as close as possible to the VISIR observations have been provided by M. Salway and A. Wesley.  Suppressed emission at both 4.8 and 8.6 $\mu$m is caused by excess aerosol opacity.   NSFCAM2 images were taken on different nights to the VISIR images, and do not show the same longitude range.  Turbulent activity is evident in rows (a) and (b) (the `wake' of the GRS), and the chain of three barges (B1-B3) are labelled in the SEB(S) in row (c).  The 2008 conjunction of Oval BA and the GRS can be seen in rows (a) and (b).  Large white arcs seen equatorward of 5\degree S in the 8.6-$\mu$m images are due to the removal of negative-beam artefacts caused by the small 20" chopping amplitude. }
\label{TBmaps1}
\end{figure*}

\begin{figure*}[tbp]
\centering
\epsfig{file=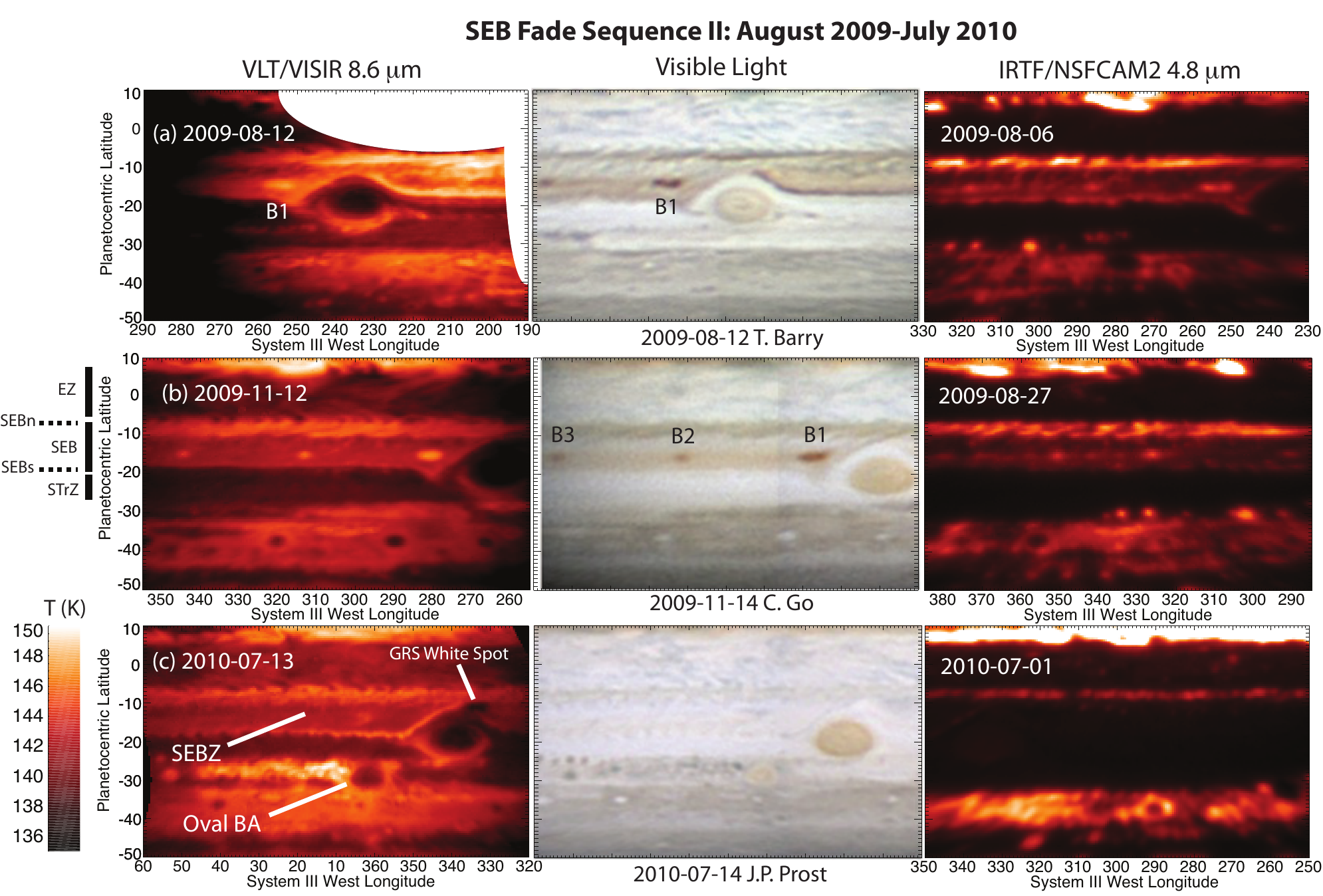,width=\textwidth,angle=0}
\caption{The SEB fade sequence August 2009-July 2010, observed by VLT/VISIR at 8.6 $\mu$m (left, sensitive to aerosol opacity at $p<800$ mbar); visible light from amateur observers (centre); and IRTF/NSFCAM2 observations at 4.8 $\mu$m (right, sensitive to aerosols above the 2-3 bar level).  Visible images taken as close as possible to the VISIR observations have been provided by T. Barry, C. Go and J.P. Prost.  Suppressed emission at both 4.8 and 8.6 $\mu$m is caused by excess aerosol opacity.   NSFCAM2 images were taken on different nights to the VISIR images, and do not show the same longitude range.  The chain of three barges (B1-B3) are labelled in the SEB(S) in row (b).  The 2010 conjunction of Oval BA and the GRS can be seen in row (c).  Large white arcs seen equatorward of 5\degree S in the 8.6-$\mu$m images of row a are due to the removal of negative-beam artefacts caused by the small 20" chopping amplitude, which was increased to 25" in November 2009. }
\label{TBmaps2}
\end{figure*}

Jupiter's axisymmetric structure, consisting of bright zones and dark brown belts, can undergo dramatic visible changes over short time scales.  The most impressive of these is the variability of the South Equatorial Belt (SEB, 7-17\degree S), which can change from the broadest and darkest belt on the planet to a white zone-like appearance over a matter of months.  The SEB, which lies in Jupiter's southern tropics and contains the Great Red Spot (GRS) at its southern edge, is typically a dark brown stripe encircling the globe between two opposing zonal flows (Fig. \ref{TBmaps1}a):  a prograde (eastward) jet at 7\degree S (SEBn) and the planet's fastest retrograde (westward) jet at 17\degree S (SEBs, all latitudes are given as planetocentric).   The SEB is a site of intense convective activity and lightning storms \citep{04ingersoll}, and is one of the few locations where spectroscopically identifiable ammonia clouds \citep[SIACs, ][]{02baines} have been observed.   However, this activity and the dark colouration of the SEB were completely absent when Jupiter emerged from behind the Sun during the 2010 apparition, replaced by the pale `faded' state (Fig. \ref{TBmaps2}c).  This missing jovian belt captured the imagination of amateur and professional astronomers alike, and it prompted a program of thermal infrared imaging of Jupiter's faded SEB from the ESO Very Large Telescope (VLT) at Cerro Paranal in Chile.  These data, along with supporting observations from the NASA Infrared Telescope Facility (IRTF) and amateur observers, will be used to determine the variations in temperature and aerosol opacity within the SEB between 2008 and 2010 and provide insights into the underlying physicochemical mechanisms responsible for these dramatic modifications to Jupiter's appearance.

The SEB fade and revival cycles appear to follow a repeatable pattern, albeit at irregular and unpredictable intervals because the underlying physical causes are unknown.  Excellent historical accounts of the SEB life cycle at visible wavelengths can be found in \citet{58peek}, \citet{95rogers} and \citet{96sanchez_jup}.  The 2009-2010 fade is the start of the fifth SEB life cycle since the first spacecraft encounter with Jupiter (Pioneer 10 in December 1973), and the first to be investigated in detail in the thermal infrared using the high spatial resolutions and broad wavelength coverage of modern telescopes and instrumentation.  Pioneer 10 and 11 visited Jupiter during the 1972-1975 faded state prior to the July 1975 revival \citep{95rogers, 81orton}.  After a 14-year hiatus, the SEB faded and revived in 1989-1990 \citep{92yanamandra, 93kuehn, 94satoh} and 1992-1993 \citep{96sanchez_SEB, 97moreno}.  The next `partial' fade began in 2007:  New Horizons observations in January-February 2007 revealed the absence of both the chaotic turbulence and fresh NH$_3$ ice clouds northwest of the GRS \citep{07reuter, 07baines}, but retrievals of cloud opacity from VLT/VISIR still demonstrated a cloud-free SEB \citep{10fletcher_grs}.   A fade had started, but a violent revival began much earlier than expected, restoring dark coloration and turbulent activity by the end of 2007 \citep{07rogers, 07rogers_climax}.   This study concerns the fading event between May 2009 and July 2010.

The general pattern of the SEB life cycle can be summarised as follows.  After the cessation of turbulent rifting and convective events to the northwest of the GRS, the SEB (7-17$^\circ$S planetocentric latitude) fades to a pale colour over a matter of months, obscuring the southern component of the SEB (SEB(S), 15-17$^\circ$S) and leaving a narrow northern component (SEB(N), 7-10$^\circ$S) which has also been observed to fade in some years.  During this unusual phase the GRS appears as a conspicuous red oval surrounded by white aerosols.  The faded state can persist for 1-3 years before a spectacular revival begins with a single, localised disturbance (the SEBD).  Vigorous eruptions generate complex patterns with bright and dark coloration throughout the SEB, encircling the planet and ultimately restoring the typical brown colour.  

The aim of this research is to reconcile physicochemical variability (temperatures, composition, clouds) derived from 7-20 $\mu$m VLT imaging with visible changes in the albedo of the SEB.  The vertical temperature structure and spatial distribution of aerosols during the fade are used to differentiate between different mechanisms for the `disappearance' of the SEB.  Section \ref{data} introduces the sources of infrared and visible data and the techniques used to determine temperatures and aerosol opacities.  Section \ref{results} presents a timeline for the 2009-2010 fade, which is used to provide insights into the underlying mechanisms for the SEB fade in Section \ref{discuss}.

\section{Observations and Analysis}
\label{data}

\subsection{VLT Imaging}

High spatial resolution imaging of Jupiter in the Q (17-20 $\mu$m) and N (8-14 $\mu$m) bands from the ESO Very Large Telescope (VLT) mid-infrared camera/spectrograph \citep[VISIR,][]{04lagage} was previously used to probe the atmospheric structure of Jupiter's Great Red Spot \citep{10fletcher_grs} and to study the aftermath of the 2009 asteroidal collision \citep{11orton, 10fletcher_cxhy}.  We were awarded four hours of Directors Discretionary Time (program 285.C-5024A) to characterise the structure and composition of Jupiter's faded SEB in July 2010, which we compare to previous VISIR observations acquired between May 2008 and November 2009 (Table \ref{tab:data}) using identical imaging techniques.  

Jupiter is too large to fit entirely within the VISIR field of view ($32"\times32$"), so northern and southern hemispheres were imaged separately, often on different dates.   Furthermore, thermal imaging requires chopping between the target and an off-source position to detect the jovian flux on top of the background telluric emission.   However, the maximum chopping amplitude of VLT/VISIR is limited to 25", meaning that some of the `sky' image is obscured by the planet, preventing the use of some regions of the chopped-differenced image (e.g., white regions removed from 8.6-$\mu$m images in Fig. \ref{TBmaps1}).  Further background stability was achieved by offsetting the telescope to a sky position 60" from Jupiter (nodding).

An imaging sequence typically featured eight filters (Table \ref{tab:filters}) sensitive to (a) atmospheric temperatures via the collision-induced hydrogen continuum and stratospheric hydrocarbon emission; and (b) tropospheric ammonia and aerosols (8.59 and 10.77 $\mu$m).  Each filter was observed twice, with a small 1-2" dither to fill in bad pixels on the detector for all eight filters.  A full imaging sequence required approximately 40 minutes.  The ESO data pipeline was used for initial reduction and bad-pixel removal via its front-end interface, GASGANO (version 2.3.0\footnote{http://www.eso.org/sci/data-processing/software/gasgano/}).  Images were geometrically registered, cylindrically reprojected and absolutely calibrated using the techniques developed in \citet{09fletcher_imaging}.  Radiometric calibration was achieved by scaling the observations to match Cassini/CIRS measurements of Jupiter's radiance acquired during the 2000 flyby \citep{04flasar_jupiter}.  Consequently we cannot investigate \textit{absolute} changes in Jupiter's infrared emission, but we can study \textit{relative} variability between different latitudinal bands \citep[see ][for a complete discussion]{09fletcher_imaging}.  Measurement errors of approximately 4-6\% for each filter were estimated from the variability of the sky background in each image.  All latitudes in this study are planetocentric, all longitudes are quoted for Jupiter's System III West \citep[using the standard IAU definition of System III rotation (1965), recently reviewed by][]{10russell}.

Examples of the VISIR imaging at 8.6 $\mu$m between May 2008 and July 2010 (where low fluxes indicate either elevated aerosol opacities or cool atmospheric temperatures) are shown in Figs. \ref{TBmaps1} and \ref{TBmaps2}.  These are compared to examples of amateur images at visible wavelengths and Jupiter's 4.8-$\mu$m emission on dates close to those of the VISIR observations obtained by NASA's Infrared Telescope Facility NSFCAM2 instrument (Table \ref{tab:nsfcam}).  Jupiter's 4.8-$\mu$m emission is attenuated by aerosols above the 2-3 bar level, so dark regions indicate locations of higher opacity.  Although the dates and spatial coverage of the mid-IR and 4.8-$\mu$m images differ, their zonal properties can be compared to provide insights into the vertical distribution of cloud opacity during the fade sequence (Section \ref{results}).



\begin{table}[htdp]
\caption{VLT/VISIR Observations of Jupiter's Belts.  UT times are approximate, as each filter sequence required approximately 45 minutes to execute.}
\begin{center}
\begin{tabular}{|c|c|c|c|}
\hline
Date & Hemisphere & Time (UT) & Programme ID \\
\hline
2010-07-13 & S & 08:10 & 285.C-5024(A) \\
2010-06-23 & N & 08:50 & 285.C-5024(A) \\
2009-11-12 & S & 01:30 & 084.C-0206(B) \\
2009-08-10 & N & 03:00 & 383.C-0161(A) \\
2009-08-11 & S & 02:30 & 383.C-0161(A) \\
2009-08-12 & S & 03:00 & 383.C-0161(A) \\
2009-07-24 & S & 03:30 & 283.C-5043(A) \\
2008-09-03 & S & 04:00 & 081.C-0137(C) \\
2008-09-01 & N & 02:00 & 081.C-0137(B) \\
2008-05-17 & S & 10:00 & 381.C-0134(A) \\
\hline
\end{tabular}
\end{center}
\label{tab:data}
\end{table}%

\begin{table*}[htdp]
\caption{VLT/VISIR Filters used in this study.  Approximate peaks of the filter contribution functions are based on \citet{09fletcher_imaging}.}
\begin{center}
\begin{tabular}{|c|c|c|c|}
\hline
Name & Wavelength ($\mu$m) & Sensitivity & Approx. Pressure (mbar) \\
\hline
J7.9 & 7.90 & Stratospheric $T(p)$ & 5 \\
PAH1 & 8.59 & Aerosols and $T(p)$ & 650\\
SIV2 & 10.77 & NH$_3$, aerosols, $T(p)$ & 400 \\
NeII\_1 & 12.27 & Stratospheric $T(p)$ and C$_2$H$_6$ & 6 \\
NeII\_2 & 13.04 & Tropospheric $T(p)$ & 460 \\
Q1 & 17.65 & Tropospheric $T(p)$ & 200 \\
Q2 & 18.72 & Tropospheric $T(p)$ & 270 \\
Q3 & 19.50 & Tropospheric $T(p)$ & 400 \\
\hline
\end{tabular}
\end{center}
\label{tab:filters}
\end{table*}%

\begin{table}[htdp]
\caption{NSFCAM2 4.8-$\mu$m data used in this study}
\begin{center}
\begin{tabular}{|c|c|c|}
\hline
Date & Time (UT) & Central Meridian \\
\hline
2008-05-27 & 13:53:39 &      334.9 \\
2008-06-16 & 13:45:26 &      103.3 \\
2008-07-10 & 13:06:44 &      96.1 \\
2008-07-27 & 12:52:05 &      224.6 \\
2008-08-07 & 07:22:50 &      145.9 \\
2008-09-24 & 06:22:15 &      133.2 \\
2009-07-20 & 14:09:23 &      27.6 \\
2009-08-06 & 10:30:41 &      296.9 \\
2009-08-18 & 13:19:15 &      46.9 \\
2009-08-27 & 13:45:42 &      338.8 \\
2010-06-24 & 15:50:41 &      319.9 \\
2010-06-30 & 14:15:45 &      85.9 \\
2010-07-01 & 15:47:43 &      292.1 \\
2010-09-03 & 08:26:22 &      305.4 \\
2010-09-05 & 10:04:33 &      306.1 \\
\hline
\end{tabular}
\end{center}
\label{tab:nsfcam}
\end{table}%

\subsection{Mid-IR Retrievals}
\label{model}

To separate the effects of temperature, aerosol opacity and composition on the VISIR filtered imaging, cylindrical radiance maps in eight filters (7.9, 8.6, 10.8, 12.3, 13.0, 17.7, 18.7 and 19.5 $\mu$m) were stacked to form rudimentary image cubes using the procedure described by \citet{09fletcher_imaging}.  The temperature structure, composition and aerosol distribution were derived from the 8-point spectra using an optimal estimation retrieval algorithm, NEMESIS \citep{08irwin}.  Sources of spectral line data; conversion of line data to $k$-distributions and Jupiter's \textit{a priori} atmospheric structure were described in detail by \citet{11fletcher_trecs}.  Two different retrieval approaches were tested: (i) a single-stage process retrieving $T(p)$, NH$_3$, C$_2$H$_6$ and aerosol opacity simultaneously, and (ii) a two-stage process deriving temperatures first (from the 7.9 $\mu$m and 13.0-19.5 $\mu$m filters) before the retrieval of composition.  Atmospheric temperatures were derived as a vertical profile defined on a grid of 80 layers between 10 bar and 0.1 mbar, whereas Cassini-derived profiles of aerosol opacity, NH$_3$ and C$_2$H$_6$ \citep{09fletcher_ph3, 07nixon} were simply scaled to reproduce the VISIR spectra.  The two-stage process was found to produce unreliable results because the competing effects of temperature and aerosol opacity were inseparable without using all 8 filters simultaneously.  Thus the single-stage retrieval was used for each of the observations in Table \ref{tab:data}.  

\begin{figure}[tbp]
\centering
\epsfig{file=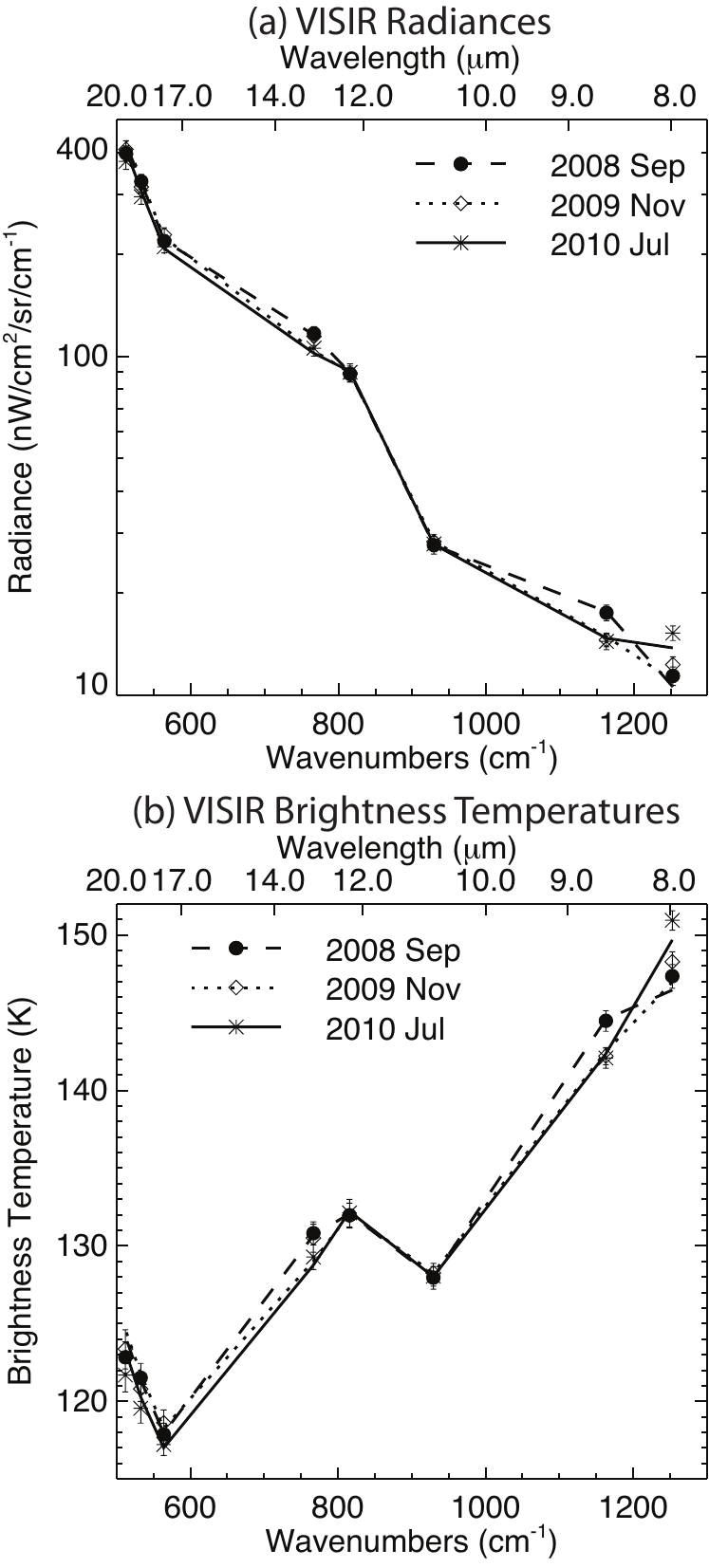,height=17cm}
\caption{Comparison of synthetic 8-point spectra (lines) to measured zonal mean radiances (points) at the centre of the SEB (14.5$^\circ$S) at three different epochs - September 2008 (normal state of the SEB); November 2009 (midway through the fade) and July 2010 (once the fade had completed).  The NEMESIS retrieval model is capable of reproducing the measured radiances in all 8 filters.  There were few differences in the calibrated radiances between the three epochs, except a cooling trend at 8.6 and 13.1 $\mu$m as the fade progressed and 7.9-$\mu$m variability due to Jupiter's quasi-quadrennial oscillation (QQO).  Radiances in the top panel were converted to brightness temperatures in the bottom panel.  The sensitivity of each of the eight filters is described in Table \ref{tab:filters}.}
\label{visirspx}
\end{figure}

Radiances from each observation in Table \ref{tab:data} were averaged over a 10$^\circ$-longitude range surrounding the central meridian, and interpolated onto a 1\degree latitude grid. Care was taken to avoid large ovals (the GRS, Oval BA) to prevent spurious results, such that central meridian scans are representative of the zonal mean radiances on each date.  Atmospheric temperatures, aerosol opacity and the distributions of tropospheric NH$_3$ and stratospheric C$_2$H$_6$ were retrieved simultaneously for each latitude to determine meridional variations in each quantity.   Examples of the quality of the 8-point spectral fits at each latitude are shown in Fig. \ref{visirspx}, which compares synthetic radiances with measurements at the centre of the SEB (12.5$^\circ$S) on three different dates.  Calibrated radiances were similar for all three epochs, with evidence of a decrease in emission at 13.0 and 8.6 $\mu$m between 2008 and 2010.  Conversely, variations of zonal mean radiances at 10.8 and 12.3 $\mu$m (sensitive to NH$_3$ and C$_2$H$_6$ respectively) were negligible, suggesting that the VISIR photometry does not show evidence for NH$_3$ or C$_2$H$_6$ variability during the fade.  It is likely that detection of gaseous variability during an SEB life cycle would require spectroscopy rather than photometry.  Nevertheless, atmospheric temperatures and aerosols could be reliably separated using the 8-point spectra and will be discussed in Section \ref{results}.

The retrieved haze opacity is driven by the 8.6-$\mu$m brightness temperatures plotted on the left of Figs. \ref{TBmaps1}-\ref{TBmaps2}, where lower brightness temperatures are the result of higher aerosol opacities.  However, the 8-point spectra lack the information content required to determine the vertical distribution of this aerosol opacity.  Cassini/CIRS spectral analyses \citep{04wong, 05matcheva, 09fletcher_ph3} demonstrated that the 8-11 $\mu$m wavelength range was best reproduced by a compact cloud layer of 10-$\mu$m radius NH$_3$ ice particles near 800 mbar, although neither the composition nor the altitude of these aerosols were particularly well constrained.  Using the CIRS-derived aerosol distribution, we scale the cumulative optical depth of the aerosols above the 800-mbar level to reproduce the 8-point spectra at each latitude.  

\section{Results:  The SEB Fade Timeline}
\label{results}

This following sections compare temperature and aerosol distributions derived from VLT 7-20 $\mu$m filtered imaging to both IRTF 4.8-$\mu$m images of deep cloud opacity and amateur imaging of the visible coloration to reveal the sequence of changes occurring during the SEB fade.  The timeline of SEB changes between 2008 and 2010 is summarised in Table \ref{tab:timeline}, along with a selection of key characteristics of the fade in Table \ref{tab:charac}.

\begin{table*}[htdp]
\caption{Timeline of events in the 2009-2010 Fade}
\begin{center}
\begin{tabular}{|l|p{10cm}|p{6cm}|}
\hline
\textbf{Date} & \textbf{Event} & \textbf{Possible Implication}  \\
\hline

2008-Sep & Typical state of SEB (Fig. \ref{TBmaps1}a): visibly brown, warm troposphere and cloud-free; turbulent activity NW of the GRS & Subsidence over much of the SEB, localised convective upwelling.\\
\hline
2009-May-26 & Turbulent convection NW of the GRS had largely ceased (Fig. \ref{TBmaps1}c) & Inhibition of discrete convective upwelling by some mechanism. \\
\hline

2009-Jun-05 & Final bright convective spots observed NW of the GRS, SEB still appears visibly brown & Cessation of SEB turbulence complete. \\
\hline

2009-Jun-15 & First dark brown cyclonic barge (B1) appeared west of the GRS \citep{10rogers} in pale brown SEB(S) (e.g., Fig. \ref{TBmaps1}c) & Barges are a new feature of the faded SEB; four more formed at progressively more westerly longitudes. \\
\hline

2009-Jul-20 & High-opacity SEBZ observed at 4.8 $\mu$m (deep clouds) separating cloud-free SEB(N) and SEB(S), visibly brown colour still present (Fig. \ref{TBmaps1}c). & Zone-like conditions beginning at depth.\\
\hline

2009-Jul-24 & Narrow lane of SEBZ higher opacity observed at 8.6-$\mu$m (upper clouds); SEB(N) and SEB(S) appear diffuse and cloud-free (Fig. \ref{TBmaps1}c) & Aerosol opacity in new SEBZ appears different in 4.8 $\mu$m (deep clouds) and 8.6 $\mu$m (upper cloud) images. \\
\hline

2009-Aug-04 & Fifth brown barge (B5) forms furthest west of the GRS & Circulation forming/revealing brown barges has propagated westward. \\
\hline

2009-Aug-06 & Opacity of SEB(S) and SEB(N) equal at 8.6-$\mu$m, but 4.8-$\mu$m opacity considerably higher over SEB(S) than SEB(N); SEB(S) 8.6-$\mu$m opacity is higher west of the GRS than to the east (Fig. \ref{TBmaps2}a); brown colour of SEB(S) fading west of GRS & SEB(S) fade west of GRS had begun.\\
\hline

2009-Oct & Brown colour of the SEB(S) east of the GRS begins to fade, bulk of the SEB acquired a pale orange tint. & Fade had progressed around the planet from its starting point west of the GRS. \\
\hline

2009-Oct-04 & Brilliant white spot observed north of the GRS \citep{10rogers} & Uncertain connection to fade process \\
\hline

2009-Nov & All barges persist in faint red-brown southern SEB, only SEB(N) remains relatively cloud-free; SEB(S) has high cloud opacity at 4.8 and 8.6 $\mu$m and cannot be distinguished from SEB (Fig. \ref{TBmaps2}a) & Fade has not yet completed, but cool SEBZ is fully established. \\
\hline

2010-Jan & End of 2009 apparition; SEB(N) narrow and dark; most of SEB pale orange; 5 barges dark and conspicuous & Fade had not yet completed. \\
\hline

2010-Apr & Start of 2010 apparition, all of SEB is pale, brown barges still faintly visible & SEB fade proceeded almost to completion while Jupiter was obscured by the Sun. \\
\hline

2010-Jul-01 & SEB(S) is completely absent in 4.8-$\mu$m imaging, SEB(N) can still be seen (Fig. \ref{TBmaps2}c).  Visible images show brown barges have almost completely faded, but still partially outlined by faint blue-grey patches. & Deep cloud opacity completely covers SEB \\
\hline

2010-Jul-13 & SEB(S) (narrow and undulating) and SEB(N) (broad) are detectable at 8.6 $\mu$m, barges cannot be distinctly observed (except at 10.8 $\mu$m, Fig. \ref{south10.8}); central SEB covered by high opacity cloud (Fig. \ref{TBmaps2}c) & SEB fade has completed.\\


\hline
\end{tabular}
\end{center}
\label{tab:timeline}
\end{table*}%

\begin{table*}[htdp]
\caption{Characteristics of note during the SEB fade}
\begin{center}
\begin{tabular}{|l|c|p{12cm}|}
\hline
\textbf{Characteristic} & \textbf{Typical Latitude} & \textbf{Temporal Behaviour}  \\
\hline

GRS Wake & 7-20$^\circ$S  & Filamentary turbulence normally dominates region NW of GRS (Fig. \ref{TBmaps1}a-b; became quiescent in May-June 2009 (Fig. \ref{TBmaps1}c); prelude to the formation of SEBZ and the fade.\\
\hline
SEB(N) & 7-12$^\circ$S & Dark, narrow brown lane separating SEB and EZ (Fig. \ref{TBmaps1}a); warm cloud-free conditions persisted throughout fade; opacity increased but did not obscure 4.8 and 8.6-$\mu$m emission completely (Fig. \ref{TBmaps2}c).  \\
\hline
SEB(S) & 15-18$^\circ$S & Broad dark band of variable width separating SEB and STrZ; began to fade west of GRS in June-July 2009; east of the GRS by October-November 2009.   By July 2010 deep cloud opacity completely obscured SEB(S) 4.8-$\mu$m emission; whereas a narrow lane of low upper-cloud opacity (8.6 $\mu$m) was visible (Fig. \ref{TBmaps2}c). \\
\hline
SEBZ & 11-15$^\circ$S & Cool zone-like temperatures developed July-August 2009 for $p>300$ mbar and persisted to July 2010 (Fig. \ref{south10.8}; Fig. \ref{tempmaps}); coincides with increased opacity of SEB centre and reduction of zonal windshear in upper troposphere (Fig. \ref{Tcompare}).\\
\hline
SEB(S) Undulations & 17-19$^\circ$S & Unique characteristic of fully-faded state in July 2010 (Fig. \ref{TBmaps2}c); zonal oscillations of cloud opacity with 5-6$^\circ$ longitude wavelength and retrograde motion; further south than the typical SEB(S) edge. \\
\hline
SEB(N) Projections & 6-9$^\circ$S & Persistent features of SEB(N), wave activity at the boundary between clouds of the EZ and SEB mixes cloud opacity.  No change during the fade.  \\
\hline
Brown Barges & 14-16$^\circ$S & Five dark brown barges in the SEB(S) (B1-B5) formed west of the GRS between June-August 2009 (as the SEBZ was forming, Fig. \ref{TBmaps2}).  All were cloud-free with warm core temperatures.  Not observed in previous fades \citep{96sanchez_jup}.  Barges faded by July 2010, but warm cyclonic circulation was still present (Fig. \ref{south10.8}).\\
\hline
GRS White Spot & 9-11$^\circ$S & White cloud (vigorous convective plume) near N edge of GRS in July 2010; high 8.6-$\mu$m opacity (Fig. \ref{aermaps}c), gives rise to a blue-grey streak northwest of the GRS \citep{10rogers}. Uncertain connection to the fade process, but may have also been present during 1989-1993 fades \citep{10rogers}.  \\
\hline
SEBD & 7-20$^\circ$S & Expected vigorous convective disturbance that will signal the start of the SEB revival.\\
\hline
\end{tabular}
\end{center}
\label{tab:charac}
\label{lasttable}
\end{table*}%

\subsection{Initial State of the SEB (Pre-May 2009)}

Filamentary turbulent convective activity usually dominates the SEB to the northwest of the Great Red Spot (GRS) in a region known as the `GRS wake.'  The chaotic activity is driven by the convergence of a system of complex atmospheric flows, where retrograde flow (from the east) at 17\degree S is deflected northward around the periphery of the GRS to meet the prograde flow (from the west) at 7\degree S.  This typical state was observed throughout 2008 (Fig. \ref{TBmaps1}a-b), where the SEB between 7-17$^\circ$S has a higher temperature than the equatorial zone (EZ) to the north and the South Tropical Zone (STrZ) to the south.  Cloud opacity measured at both 4.8 and 8.6 $\mu$m is typically low, so that the SEB appears relatively bright at both of these wavelengths.   This is consistent with the typical view of upper tropospheric belt-zone circulation \citep[e.g.,][]{04ingersoll} whereby air rises at the equator and subsides over the neighbouring belts, creating warm, cloud-free and volatile depleted conditions within the SEB and NEB (North Equatorial Belt).  Small white spots and rifting observed in the visible in Fig. \ref{TBmaps1}a-b coincide with higher aerosol opacities at 8.6 $\mu$m and cooler tropospheric temperatures, indicating localised upwelling within the generally subsiding belt.  Such upwelling transports spectroscopically identifiable NH$_3$ ice \citep[SIACs,][]{02baines} upwards into the GRS wake region.   

Table \ref{tab:timeline} shows that this SEB activity continued until May-June 2009, when amateur imaging revealed the absence of any convective white spots \citep{10rogers},  just prior to the first mid-IR images in July 2009 (Fig. \ref{TBmaps1}c).  Although the cause of this inhibition of convective activity cannot be identified in the time sequence presented here, it signalled the start of a remarkable transformation in the SEB over the following 12 months.


\subsection{Formation of the SEBZ (July-August 2009):}

A comparison of the visible, 8.6 and 4.8-$\mu$m images in July 2009 (Fig. \ref{TBmaps1}c) shows a substantial alteration to the distribution of aerosol opacity within the SEB at a time when the visible colours of the SEB were largely unaltered.  At this time, the SEB had a pale interior separating a narrow brown northern component (the SEB(N), 7-10$^\circ$S) and a broad brown southern component (the SEB(S), 14-18$^\circ$S).  Because there is minimal gas opacity at 4.8 $\mu$m, strong thermal emission implies a relative dearth of aerosol opacity above the 2-3 bar pressure level \citep[e.g.,][]{98roos-serote}; whereas 8.6-$\mu$m is sensitive only to clouds and hazes above the 800-mbar level.  An enhancement in cloud opacity above the 2-3 bar level in June-July 2009 caused a substantial reduction in 4.8-$\mu$m emission from the SEB, restricting cloud-free conditions to the SEB(N) and SEB(S).  The transformation at 8.6 $\mu$m was equally dramatic - the fine structures that characterised the VISIR 8.6-$\mu$m imaging in 2008 had been replaced by a more diffuse appearance, with a narrow dark lane (8-13$^\circ$S) of elevated opacity at the 800-mbar level.  This lane of elevated aerosol opacity is known as the SEB zone (SEBZ).

\begin{figure*}[tbp]
\centering
\epsfig{file=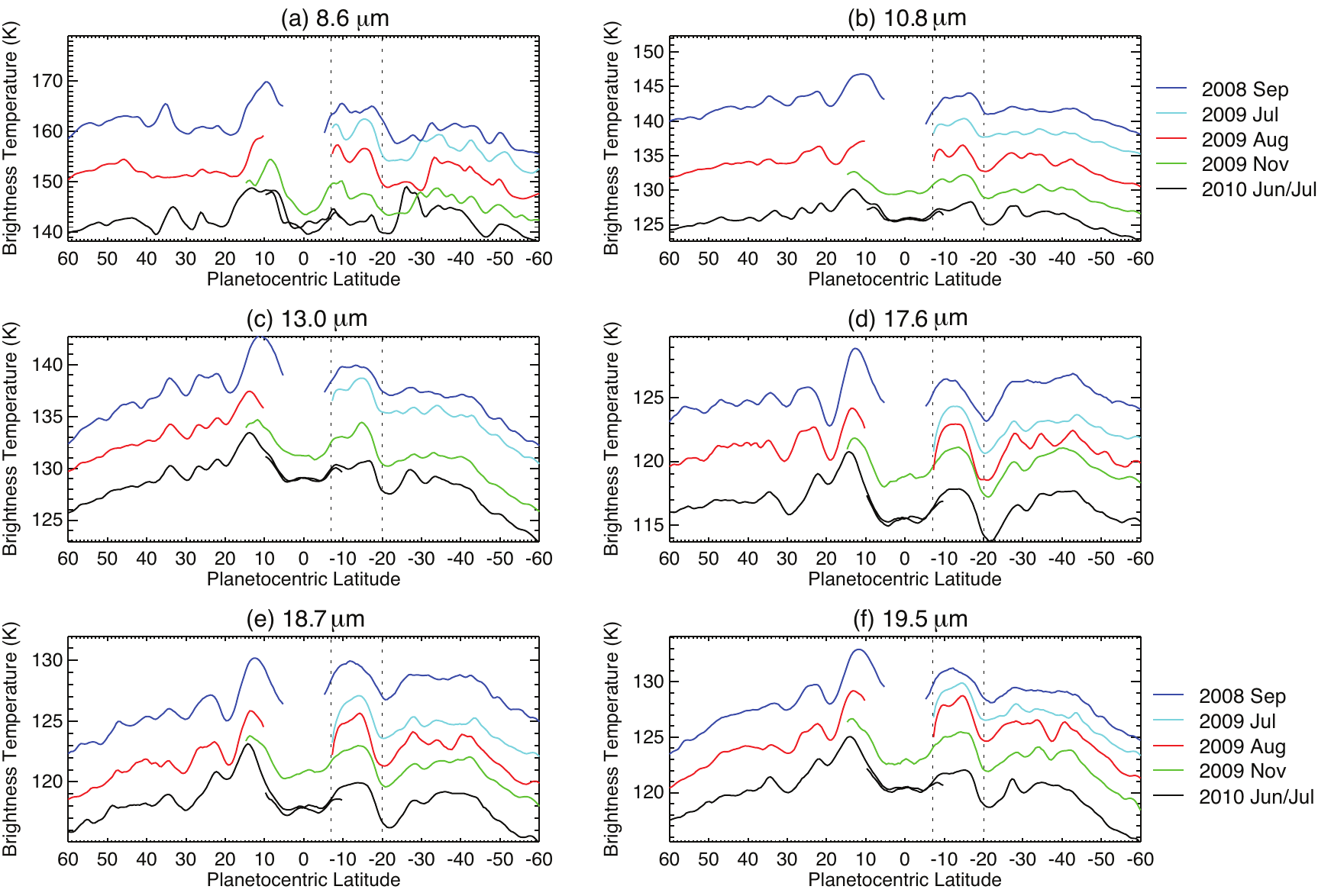,width=\textwidth}
\caption{Central meridian brightness temperature scans from 2008 to 2010 in six of the eight filters observed by VISIR.  Northern and southern hemisphere images were not acquired simultaneously, so a gap is present at the equator for the earliest dates (Table \ref{tab:data}).  The location of the SEB is denoted by vertical dotted lines.  The brightness temperatures have been offset from the 2010 measurements (black curve) by arbitrary amounts for clarity.  The most dramatic change occured at 8.6 $\mu$m (decreased emission due to the increased aerosol opacity over the SEB).  The formation of the cool SEBZ is also apparent in filters sensitive to temperatures at $p>300$ mbar (10.8, 13.0 and 19.5 $\mu$m), but only a subtle `flattening' of the meridional temperatures can be observed at 17.6 and 18.7 $\mu$m (sensitive to 150-300 mbar).  Brightness temperatures have not been corrected for emission angle, so the general decrease from equator to pole in each filter is the effect of limb darkening. }
\label{cmerid_TB}
\end{figure*}

\begin{figure}[tbp]
\centering
\epsfig{file=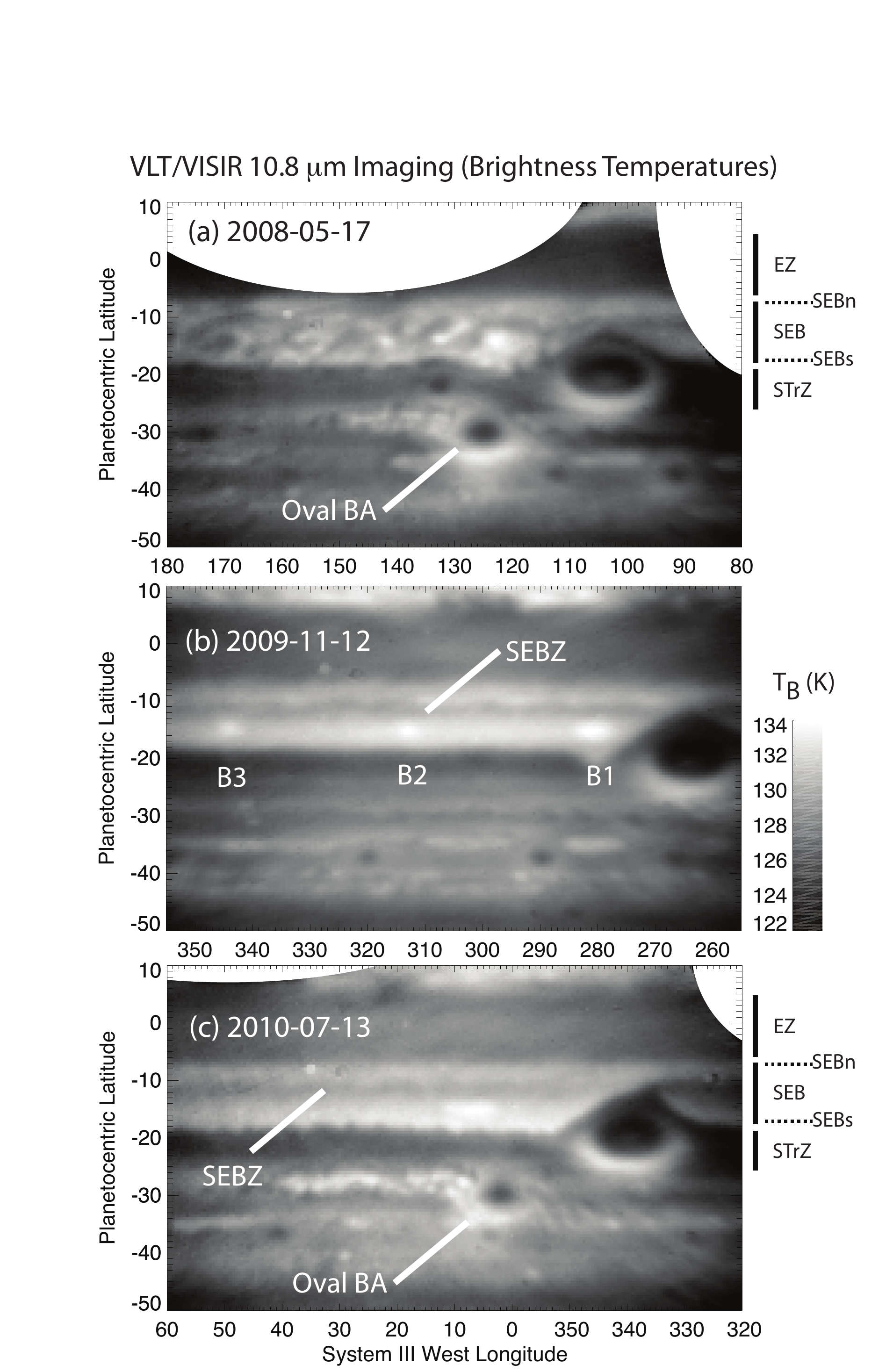,height=15cm}
\caption{Examples of Jupiter's 10.8-$\mu$m emission (sensitive to tropospheric temperatures near 600 mbar and ammonia gas abundance) from the southern hemisphere during the 2008-2010 fade.  The SEB(S) appears brighter than the northern component in 2009 and 2010, and the dark SEBZ becomes more prominent as the fade progresses (panels b and c).  Large white arcs seen equatorward of 5\degree S in panel (a) and (b) are due to the removal of negative-beam artefacts caused by the small 20" chopping amplitude.  The location of the warm brown barges and Oval BA are indicated. }
\label{south10.8}
\end{figure}

Fig. \ref{cmerid_TB} shows how the establishment of the SEBZ modified central-meridian brightness temperatures across the SEB between 2008 and 2010.  These brightness temperatures are averaged within 10$^\circ$ longitude of the central meridian for each filter, so are not true zonal averages.  Hence, some of the apparent variability results from the presence of discrete features (waves, vortices) close to the central meridian.  Changes associated with the SEB fade between 7 and 20\degree S are most dramatic at 8.6 $\mu$m, with the whole region becoming darker in 2009-2010 than the south temperate region poleward of 25\degree S.  Variations at other wavelengths are more subtle:  filters with contribution functions sensitive to upper tropospheric temperatures (150-300 mbar, approximately, for 17.6 and 18.7 $\mu$m, Table \ref{tab:filters}) remained largely unchanged throughout the fade sequence when compared to mid-IR variability at other latitudes.  Conversely, close inspection of images probing higher pressures ($p>300$ mbar: 8.6, 10.8, 13.0 and 19.5 $\mu$m) demonstrate the development of a multi-peaked structure, with radiance maxima at 9\degree S (SEB(N)) and 15-17\degree S (SEB(S)), and a zone-like minimum at 12\degree S (the SEBZ).  The transition from the turbulent SEB to the cool SEBZ by late 2009 is clearly shown in maps of 10.8-$\mu$m brightness temperatures (Fig. \ref{south10.8}), sensitive to temperatures and ammonia near the 400-mbar level. 

\subsubsection{SEBZ Temperatures}

\begin{figure*}[tbp]
\centering
\epsfig{file=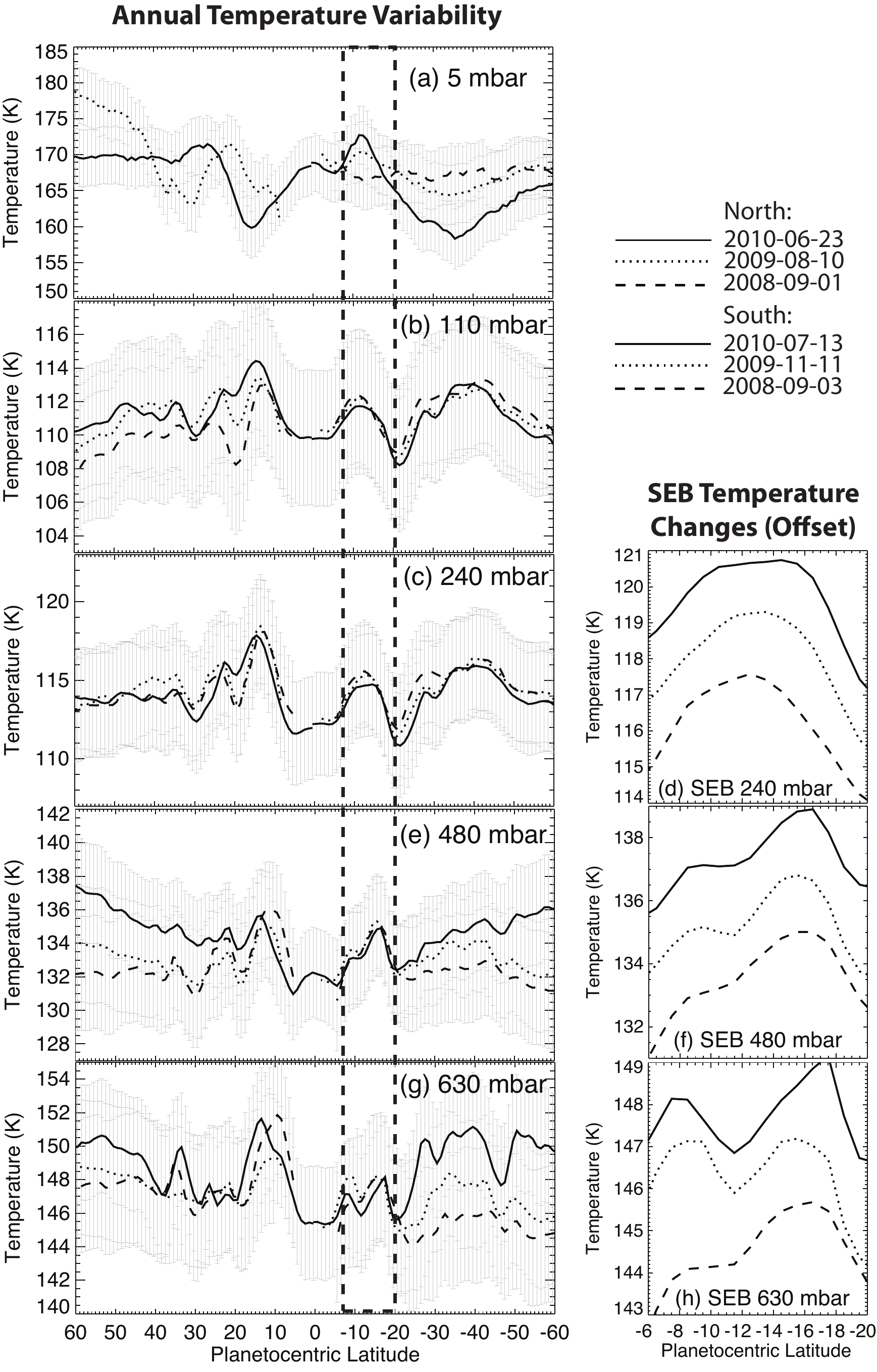,width=14cm}
\caption{\textit{Left:}  Comparing retrieved meridional temperatures at five different pressure levels for 2008 (dashed line), 2009 (dotted line) and 2010 (solid line).  The location of the SEB is denoted by vertical dashed lines.  The grey bars represent the formal uncertainty on the retrieved quantities at each altitude (i.e., taking correlations between parameters into account), which appear large because of the limited retrieval capabilities from 8-point spectral retrievals and the uncertainties in absolute calibration.   However, relative variability in the quantities are robust (similar variations are seen in each filter), even if the absolute values are uncertain.  Some disconnect between northern and southern hemisphere results is apparent at the equator because these observations were acquired on different dates, as shown by the key (top right).  \textit{Right:} Close-up views of thermal changes associated with the SEB at 240, 480 and 630 mbar, with offsets of 2 K from the 2008 temperature profiles to highlight differences.}
\label{Tcompare}
\end{figure*}

The changes in the mid-IR radiance in Fig. \ref{cmerid_TB} manifest themselves as variations in the retrieved tropospheric temperatures in Fig. \ref{Tcompare}, particularly those at high pressures (480 and 630 mbar in the panels f and h), indicating a difference in the latitudinal temperature gradient between 2008 and 2009-10.  Uncertainties in the absolute calibration of the VISIR images, combined with the broad pressure range covered by the weighting function for each filter \citep{09fletcher_imaging} and the degeneracies between temperature and composition, lead to considerable retrieval uncertainties for temperatures at $p>500$ mbar in Fig. \ref{Tcompare}e-h, so that \textit{absolute} cooling is difficult to detect at the SEB latitude.  On the other hand, the presence of the distinct cool zone (SEBZ) from 2009 onwards, \textit{relative} to the warmer SEB(N) and SEB(S), can be seen at 480 and 630 mbar in Figs. \ref{Tcompare}f-h.   Temperature contrasts of $1.0\pm0.5$ K between the SEBZ and the northern and southern components were measured at 630 mbar in July 2010 (Fig. \ref{Tcompare}h).  Of the three Q-band filters in Fig. \ref{cmerid_TB}, only the deep-sensing 19.5-$\mu$m filter (which has the lowest diffraction-limited resolution of all the images used in this study) detected the SEBZ, confirming that the SEBZ formation occurred at depth. The cool SEBZ was not observed at lower pressures (150-300 mbar) sensed by 17.6- and 18.7-$\mu$m filters where the SEB retained its usual warm belt-like conditions.   

Indeed, the temperature fluctuations in the stable upper troposphere between 15-300 mbar were small compared to those at other latitudes, particularly those associated with the NEB (see \ref{phenom}).  The SEBZ formation at depth did have a subtle effect at these lower pressures - both the retrieved temperatures near 240 mbar (Fig. \ref{Tcompare}d) and the raw radiances (Fig. \ref{cmerid_TB}) demonstrate a `flattening' of the meridional temperature gradient ($dT/dy$) between 7-20\degree S as the fade progressed.  Instead of having a belt with a peak temperature in the centre (12$^\circ$S), the 240-mbar temperatures during the faded state became more homogenised with latitude ($dT/dy$ tends to zero between 10-15$^\circ$S at 240 mbar, Fig. \ref{Tcompare}d).  As vertical shears on the zonal wind ($du/dz$) are related to $dT/dy$ by the geostrophic thermal windshear equation, this reduction of $dT/dy$ in 2009-10 compared to the normal state of the SEB suggests a reduction in the windshear on both the prograde SEBn jet at 7\degree S and the retrograde SEBs jet at 17\degree S in the troposphere.  Zonal flow associated with these two opposing jets could therefore persist to higher altitudes (i.e., near the tropopause) during a faded state than during the `normal' state.  

Finally, stratospheric temperatures above the SEB show considerable variability between 2008 and 2010 (Fig. \ref{Tcompare}a) as part of Jupiter's Quasi Quadrennial Oscillation \citep[QQO,][]{91leovy}, which affects the stratospheric temperatures between 20\degree N and 20\degree S with a period of 4-5 years.   However, no causal connection between the fades and the stratospheric oscillation is found:  (i) the regularity of the QQO means that it cannot be responsible for the SEB fades when some are separated by 14 years or more; and (ii) the 2009-10 observations show the stratospheric SEB warmer than the EZ, whereas data in 1989-90 and 1992-93 \citep{99friedson} show the reverse to be true. We conclude that the SEB fade had no effect on stratospheric temperatures near 5 mbar.

\subsubsection{SEBZ Aerosols}
\label{sebz_aer}

\begin{figure*}[tbp]
\centering
\epsfig{file=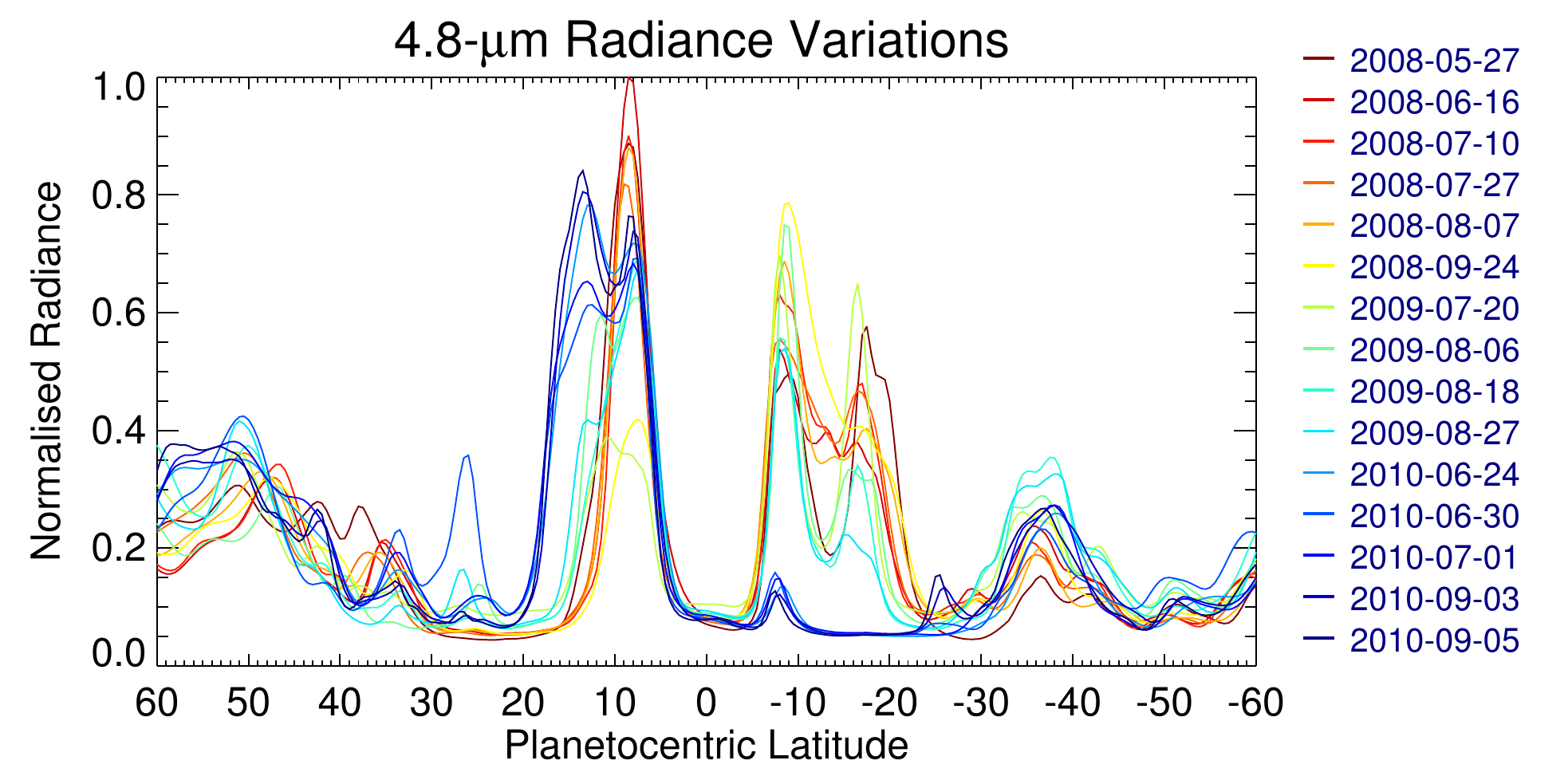,width=15cm}
\caption{Central meridian radiance scans at 4.8 $\mu$m from IRTF/NSFCAM2 (Table \ref{tab:nsfcam}) from May 2008 to September 2010.  All radiances have been normalised to the equator ($\pm5^\circ$), and then the maximum has been normalised to 1.0.  A large asymmetry between the NEB and SEB develops between 2008 and 2010.}
\label{5um_merid}
\end{figure*}

Table \ref{tab:timeline} indicates that changes to the infrared aerosol opacity at 4.8 and 8.6 $\mu$m in July-August 2009, as well as the formation of the cool SEBZ at $p>300$ mbar, occurred several months before the greatest degree of visible whitening of the SEB was observed in April 2010.  Furthermore, opacity variations at 4.8 and 8.6 $\mu$m (Figs. \ref{TBmaps1}-\ref{TBmaps2}) were morphologically different and proceeded at different rates.  Variations in cloud opacity above the 2-3 bar level are shown in the latitudinal variations of 4.8-$\mu$m emission (Fig. \ref{5um_merid}) extracted from NSFCAM2 images (Table \ref{tab:nsfcam}).  Radiances were extracted within 10$^\circ$ of the central meridian for each image and have been normalised to the equator ($\pm5^\circ$) for ease of comparison.   Between 2008 and 2010 an extreme asymmetry developed between the SEB and NEB (which was also undergoing changes in emission, see \ref{phenom}), with a notable change in contrast between the bright SEB(N) and the dimmer SEB(S) occurring between July and August 2009 (Figs. \ref{TBmaps1}-\ref{TBmaps2}), coincident with the first VLT observations of the SEBZ.    The 4.8-$\mu$m radiance of the SEB(S) at 16-18$^\circ$S diminished steadily from July-August 2009 onwards.  By the time of the first NSFCAM2 observations in June 2010, by which time the SEB visible whitening was complete, the SEB(S) emission had vanished in its entirety and the SEB(N) emission was substantially reduced, a state that persisted until at least September 2010 (the final NSFCAM2 observations of this study).

\begin{figure}[tbp]
\centering
\epsfig{file=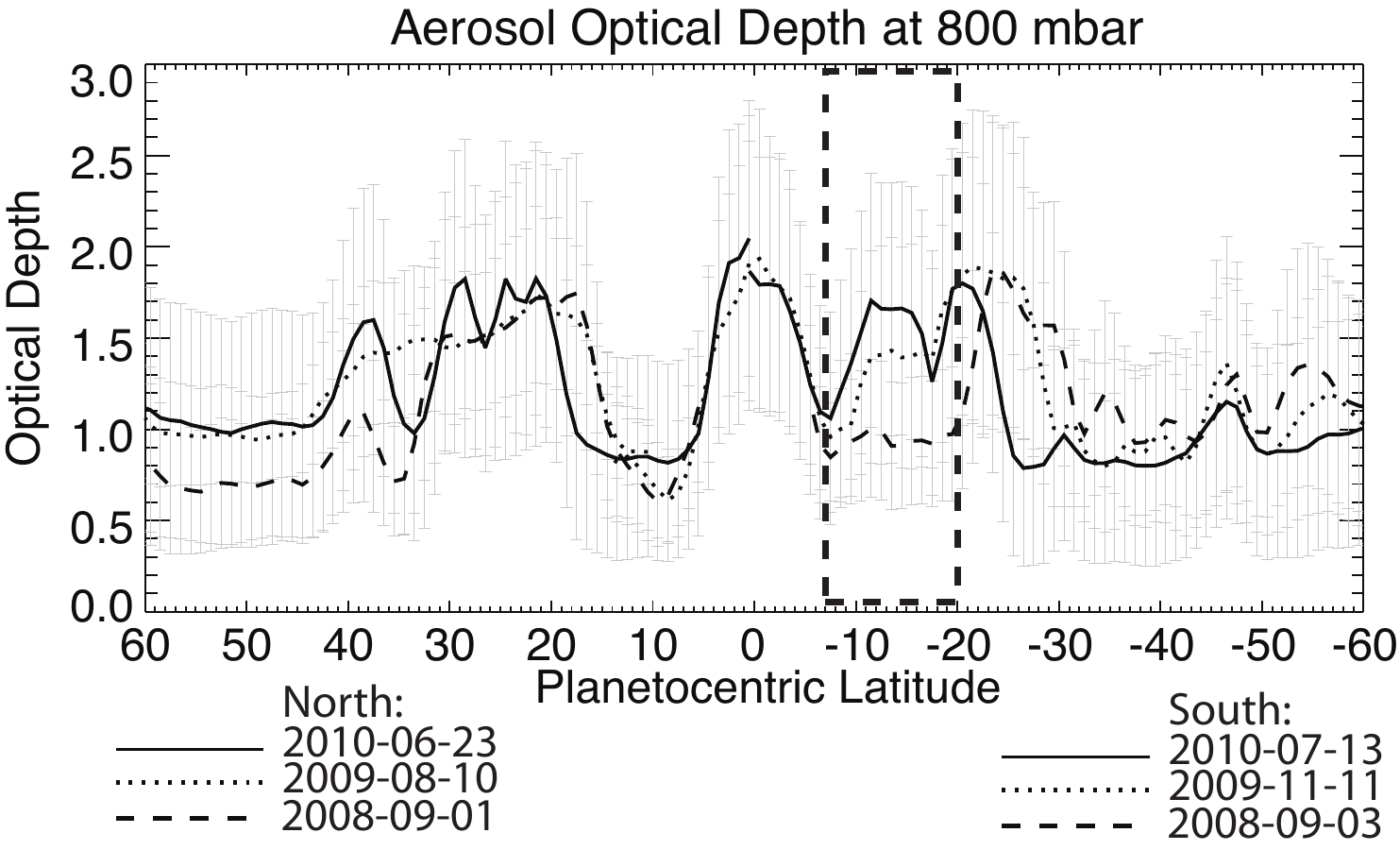,width=8cm}
\caption{Comparison of spatial distributions of aerosol opacity for 2008 (dashed line), 2009 (dotted line) and 2010 (solid line).  Opacities retrieved from 8-point 7-20 $\mu$m imaging are sensitive to aerosols above the 800-mbar level, tentatively identified as an NH$_3$ ice cloud \citep[e.g.,][]{04wong, 05matcheva}. The grey bars represent the formal uncertainty on the retrieved quantities at each altitude.  Some disconnect between northern and southern hemisphere results is apparent at the equator because these observations were acquired on different dates.  The SEB is highlighted by the dashed-line box which highlights the development of an asymmetry between the NEB and SEB between 2008 and 2010.}
\label{gas_compare}
\end{figure}

Opacity variations at lower pressures ($p<800$ mbar) sensed by VISIR were separated from thermal variations by the NEMESIS retrievals described in Section \ref{model}.     Fig. \ref{gas_compare} shows the latitudinal variation of the cumulative opacity above the 800-mbar level, with the SEB region highlighted by a dashed box.  Opacities of the SEB and NEB were similar in 2008 (a factor of 2 smaller than the equatorial opacity), but the SEB opacity then increased dramatically between 2009 and 2010.  By July 2010, the SEB opacity was 90\% of that in the EZ (Equatorial Zone), compared to 50\% in 2008.  This was caused by an increased opacity of 80\% compared to September 2008.  The dramatic `clouding-over' of the SEB is most apparent in Fig. \ref{aermaps}, where aerosol opacity was retrieved from 8-point spectra on a $0.5\times0.5^\circ$ grid.  Retrieved aerosol opacities in September 2008 showed the GRS as an isolated vortex of high aerosol opacity (Fig. \ref{aermaps}a), with cloud-free conditions over the SEB caused by wide-spread subsidence \citep[similar cloud-free conditions were previously identified by][throughout much of the 2000-2008 period]{10fletcher_grs}.  But by November 2009 the clouds over the SEB had become distinctly more opaque; and by July 2010 the majority of the SEB interior had the same opacity as the STrZ (South Tropical Zone), separated from the adjacent bands by a narrow cloud-free SEB(S) and wider SEB(N) (Fig. \ref{aermaps}c).

\begin{figure*}[tbp]
\centering
\epsfig{file=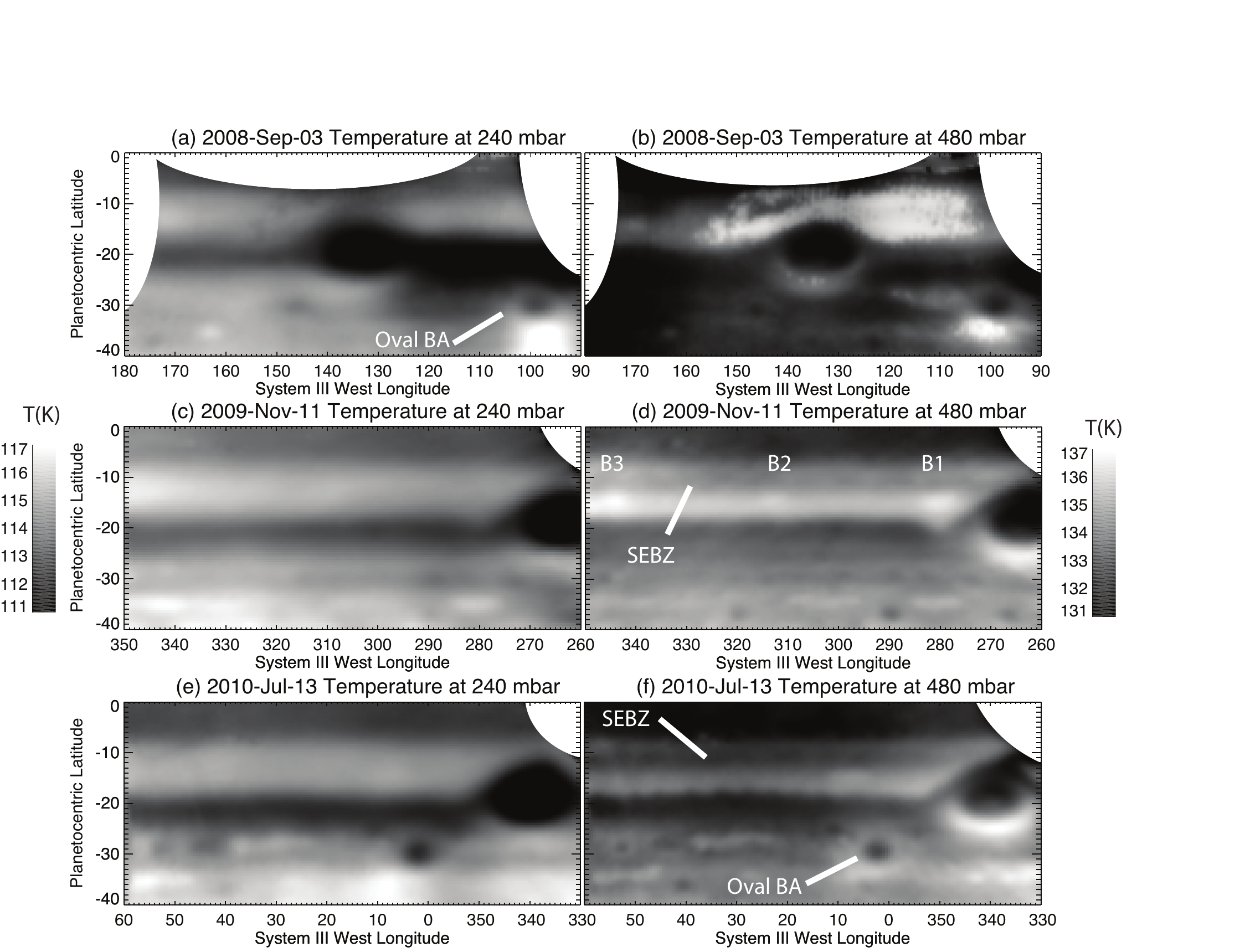,width=\textwidth,angle=0}
\caption{Retrieved temperature maps at 240 mbar (left) and 480 mbar (right).  White ellipses mark regions contaminated by artefacts due to (i) the small chopping amplitude used by VLT and (ii) cylindrical reprojections of the images.  The cool SEBZ is present in 2009 and 2010 in the 480-mbar maps, but cannot be seen at 240 mbar.  The vertical variation of zonal mean temperatures is shown in Fig. \ref{Tcompare}.}
\label{tempmaps}
\end{figure*}

\begin{figure}[tbp]
\centering
\epsfig{file=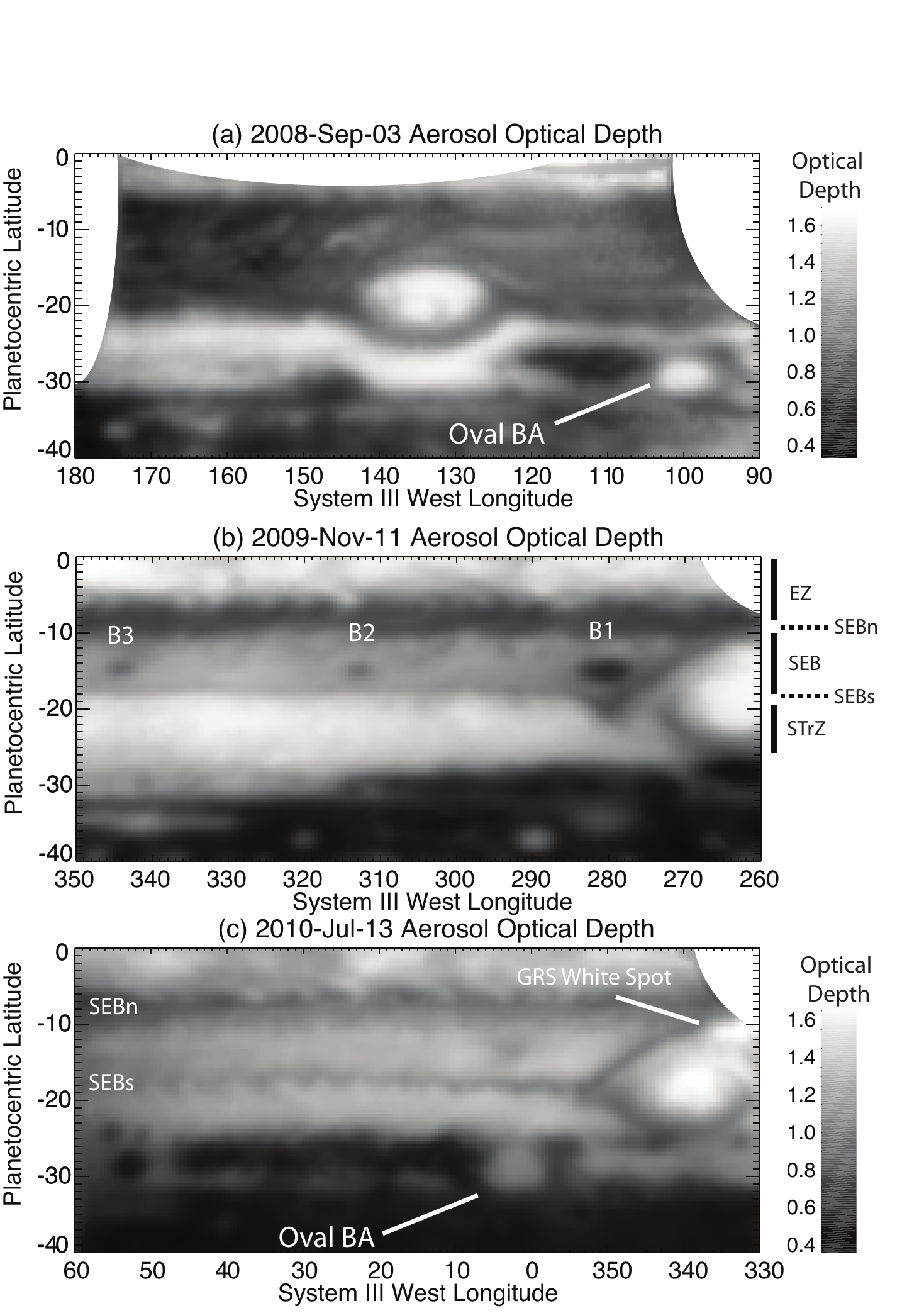,height=13cm}
\caption{Retrieved maps of aerosol opacity for clouds above the 800-mbar pressure level as the SEB faded.  The retrieved optical depth was modelled as the cumulative infrared opacity of 10-$\mu$m radius NH$_3$-ice particles with a base at 800 mbar, although NH$_3$ has not been spectroscopically identified (see Section \ref{model}).  The increase in aerosol opacity over the SEB is dramatic, with the SEB having the same opacity as the Equatorial Zone (EZ) and the South Tropical Zone (STrZ) by July 2010.  Aerosols within the GRS remain separated from these white aerosols by a cloud-free ring.   White ellipses mark regions contaminated by artefacts due to (i) the small chopping amplitude used by VLT and (ii) cylindrical reprojections of the images. }
\label{aermaps}
\end{figure}

The SEB(N) (7-10$^\circ$ in Fig. \ref{gas_compare}) remained cloud-free throughout the fade, with only a small rise in opacity consistent with the detection of SEB(N) 4.8-$\mu$m emission (albeit attenuated) in 2010 (Fig. \ref{5um_merid}).  The SEB(S) was observed as a diffuse band at 8.6 $\mu$m in July-August 2009 (Fig. \ref{TBmaps2}a), but its opacity had increased dramatically by November 2009 (Fig. \ref{TBmaps2}b).  This transition at 8.6 $\mu$m coincided with the initial visible fading of the SEB(S) and the disappearance of 4.8-$\mu$m emission (Table \ref{tab:timeline}).  Finally, although Figs. \ref{TBmaps2}c and \ref{aermaps}c suggest that the SEB(S) opacity had decreased again in July 2010 (it can be seen as a narrow, bright 8.6-$\mu$m streak west of the GRS), this was almost certainly a temporary event associated with a narrow, visibly-dark streak emanating from the bright plume north of the GRS (visible in Fig. \ref{TBmaps2}c as a spot of high 8.6-$\mu$m opacity).

The differences between 4.8- and 8.6-$\mu$m aerosol opacity are evident in both July 2009 and July 2010.  At the start of the fade in 2009 (Fig. \ref{TBmaps1}c) 4.8-$\mu$m opacity was restricted to narrow bands in the SEB(N) and SEB(S), whereas 8.6-$\mu$m opacity was much broader.  By the completion of the fade in July 2010 (Fig. \ref{TBmaps2}c), 4.8-$\mu$m emission from the SEB(S) was undetectable, whereas it could be faintly distinguished at 8.6 $\mu$m.  The attenuation of 4.8-$\mu$m flux cannot be solely explained by the increased opacity retrieved from the 8.6-$\mu$m filter.  These differences suggest sensitivity to two different populations of cloud and aerosol particles: an upper cloud with greater transparency near the SEB(N) and SEB(S), and a deeper cloud with opacity completely obscuring the bulk of the SEB and the SEB(S).  This implies that the circulation regimes associated with the faded SEB varied with height, with subsidence causing cloud-clearing over the SEB(S) at $p<800$ mbar but not at $p>800$ mbar.  Modelling of simultaneous 4.8-20 $\mu$m spectroscopy (at low to moderate resolutions) will be required to separate these different vertical regimes.  

\subsection{Lifetime of the Brown Barges (June 2009-July 2010)}

Between June and August 2009, a series of five dark brown cyclonic ovals (`barges', B1-B5) formed to the west of the GRS \citep{10rogers} at 15-16\degree S (Figs. \ref{TBmaps1}-\ref{TBmaps2}).   They formed within the broad SEB(S), which was still visible as a pale orange band between 14-18$^\circ$S during this interval and displayed warmer temperatures than the SEBZ to the north (Fig. \ref{tempmaps}d).  The first barge, B1, appeared on June 15, just west of the GRS (visible in Fig. \ref{TBmaps1}c).  Four others then appeared at progressively higher longitudes, with B5 the last to appear on August 4.  These barges may be manifestations of a deeper underlying circulation, and had not been visible in the preceding convective state of the SEB.  They have not been frequently observed during previous SEB fades \citep[summarised by][]{96sanchez_jup}, but the 2007 revival (following the partial fade) may have started with a convective plume within one such barge \citep{07rogers_climax}, suggesting an important role for these barges in the upcoming revival.  The barges were bright at both 4.8 and 8.6 $\mu$m, and the retrieved aerosol opacity above the 800-mbar level (Fig. \ref{aermaps}) suggests that they were isolated cloud-free regions of strong atmospheric subsidence embedded within the uniformly-opaque SEB(S).  This atmospheric subsidence warms the air within the barges, making them warmer than the surrounding SEB(S) (Fig. \ref{south10.8}, sensitive to temperatures near 400 mbar, and retrieved temperature maps in Fig. \ref{tempmaps}).  The barges were not visible in Q-band imaging sensitive to lower pressures, and do not perturb temperatures in the 240-mbar region (Fig. \ref{tempmaps}c).

When Jupiter reappeared from behind the Sun in 2010 amateur observers reported the continued presence of these barges \citep{10rogers}.  By July 2010 (Table \ref{tab:timeline}) the barges had also faded, replaced by faint blue-grey spots that marked the northeast corners of their original locations \citep{10rogers_spots}.  Fig. \ref{south10.8}c shows that the underlying warm cyclonic circulation of barge B1 was still present on July 13th, even though only a diffuse remnant of the barge could be detected at 8.6 $\mu$m (Fig. \ref{TBmaps2}c).  No evidence for the barges was detected at 4.8 $\mu$m (i.e., the radiance was completely attenuated due to aerosol opacity), and B1 could not be clearly identified in the aerosol map in Fig. \ref{aermaps}c, so we can conclude that although the barges had also faded by July 2010, the original circulation may persist below the visible clouds.  

\subsection{Completion of the Fade (October 2009-July 2010)}
%
By April 2010 the opacity of the SEB had increased dramatically (Fig. \ref{aermaps}) and the belt was very pale in visible light.  The SEB(N) had narrowed between November 2009 and July 2010 but remained present \citep[similar to the faded state of 1992,][]{96sanchez_SEB}, and could be identified as a bright band at 4.8 $\mu$m (Fig. \ref{TBmaps2}c).  The SEB(S) had a bright southern edge at 8.6 $\mu$m but was completely attenuated at 4.8 $\mu$m, and although the warm cyclonic circulation of the barges remained detectable, they produced no colour contrasts in visible light except for the faint blue-grey spots described above.

The faded SEB showed no discrete convective activity at any wavelength, although the narrow lanes in the SEB(N) and SEB(S) exhibited small-scale zonal wave activity to the west of the GRS with a longitudinal wavelength of 5-6\degree.  The SEB(N) projections (wavy edges or `chevrons' in Figs. \ref{TBmaps2}c) are typical features of this prograde SEBn jet at 6-8$^\circ$S, whereas the SEB(S) zonal wave activity at 18-20$^\circ$S is a new feature of the faded state of the belt.  This undulating appearance of the SEB(S) was also imaged during the completely faded state in 1974 \citep[Pioneer-11 images,][]{77rogers} and 1993 \citep{96sanchez_SEB}; they were not present during the partial fade of 2007 when the SEB(S) edge was completely straight, and they were not visible in the early stages of the 2009/10 fade (i.e., before October 2009).    The SEB(S) undulations were further south than the usual SEB(S) boundary (see Table \ref{tab:charac}).  Windspeed measurements for the SEB(S) structures from amateur imaging \citep{10rogers_spots} indicate that their retrograde velocities were consistent with spacecraft-derived zonal velocity measured for 19$^\circ$S (i.e., slower than the retrograde SEBs jet peak at 17$^\circ$, which is usually thought to bound the SEB(S)).   

At the time of writing (October 2010) amateurs and professionals are continuing to observe Jupiter for signs of a revival of the turbulence and the typical brown appearance of the SEB.  In Section \ref{discuss} we discuss hypotheses for the onset of the SEB fade and the possible thermal-infrared characteristics of the upcoming revival.


\section{Discussion}
\label{discuss}

Retrievals of atmospheric temperature and aerosol opacity from VLT/VISIR imaging have revealed that the increased SEB reflectivity during a fade is accompanied by (a) the formation of a cool SEBZ in the centre of the typically warm belt for $p>300$ mbar, separating the northern and southern components and decreasing the thermal windshear in this region; and (b) a dramatic increase in the opacity of jovian aerosols in the SEB in two vertical regimes:  aerosols in a cloud deck near 800 mbar attenuating 8.6-$\mu$m radiance, and clouds above the 2-3 bar level attenuating 4.8-$\mu$m radiance.  By the completion of the fade in July 2010, enhanced aerosol opacity obscured the bulk of the SEB and the SEB(S), whereas the SEB(N) was still visible and relatively cloud-free \citep[typical of previous SEB life cycles,][]{96sanchez_jup}.  The retrieved temperature distributions and the differences between aerosol distributions mapped by the 4.8- and 8.6-$\mu$m filters confirm that the SEB fade was caused by changes at depth ($p>300$ mbar), most likely within the convective region of the troposphere rather than the stable, radiatively cooled upper troposphere.  This deep change is supported by a qualitative inspection of amateur CH$_4$-band images (0.89 $\mu$m) in 2010, which indicated that the upper tropospheric hazes (100-200 mbar) remained unaffected by the fading process (A. S\'{a}nchez-Lavega, \textit{pers. comm.}). 

Thermal-IR observations between the cessation of SEB convective activity in May 2009 and the onset of the fade are crucial to distinguish between two scenarios proposed to explain the formation of the cool, optically-thick SEBZ.  Firstly, \citet{96sanchez_jup} suggested that the increased reflectivity of the opaque cloudy layer at visible wavelengths would lead to heating below the cloud (a `greenhouse effect' driven by internal radiation `trapped' by the high opacity of the cloudy layer) and cooling above, increasing the atmospheric lapse rate $\Gamma=-dT/dz$.  In this scenario, the cool SEBZ would begin to form \textit{after} the development of the thicker reflective clouds.  However, the cool SEBZ (Fig. \ref{cmerid_TB}) and the thickening infrared aerosol opacity (Figs. \ref{5um_merid} and \ref{aermaps}) were observed in 2009 \textit{before} the complete whitening of the SEB by April 2010, consistent with thermal changes \textit{causing} the fade rather than being a by-product.  

In a second scenario, the enhanced opacity of the cool SEBZ can be directly compared to conditions in Jupiter's zones, where upwelling and atmospheric divergence brings condensibles like NH$_3$ and H$_2$S upward from deeper levels to form clouds of NH$_3$ and NH$_4$SH ice \citep[i.e., the SEB takes on zone-like properties,][]{94satoh}.  Enrichments in volatiles and/or decreased tropospheric temperatures can both increase the density of an ice cloud and move it to deeper pressures \citep[e.g.,][]{73weidenschilling, 99atreya}.  This scenario suggests that zone-like upwelling forms the cool SEBZ \textit{before} the visible brightening, injecting fresh volatiles to form or reinforce the condensation cloud decks over the SEB.  The most likely candidate for the white aerosols of the faded state is NH$_3$ ice, which is expected to form the 800-mbar clouds attenuating 8.6-$\mu$m radiance \citep[e.g.,][]{73weidenschilling}.  By April 2010 the fresh ice coating had masked the blue-absorbing chromophore responsible for the belt's typical red-brown coloration, without affecting the properties of the upper-level hazes ($p<300$ mbar).   This is consistent with the analyses of visible reflectivity changes during the 1989-93 SEB fades \citep{93kuehn, 92satoh, 94satoh, 97moreno}, which revealed that both the optical thickness and the single scattering albedo of a 400-700 mbar cloud layer increased during the fade. The authors suggested that this reinforcement of the cloud deck was been driven by a fresh supply of NH$_3$ gas \citep[from upward diffusion or convective overshooting][]{93kuehn}, condensing on available nucleation sites in this region.  

The 2009-10 observations show that the SEB life cycle also affects cloud condensation at pressures greater than 800 mbar, as condensation of NH$_3$ ice alone cannot explain the different spatial distributions of aerosol opacity at 4.8 and 8.6 $\mu$m.  Other condensibles such as NH$_4$SH may be responsible for the increased opacity above the 2-3 bar level sensed at 4.8 $\mu$m, but neither of these ices have been identified spectroscopically during the present faded state of the SEB - a key challenge for future observations. 

Studies of temperature changes associated with previous SEB fades have been inconclusive - \citet{92yanamandra} reported a general SEB cooling prior to the 1989 fade from IRTF observations at 18 $\mu$m, but the published survey of \citet{94orton} concluded that the 1989-1993 fades had no contemporaneous temperature changes at 250-mbar that were distinguishable from seasonal variations (their Fig. 4).  Given that the 2009-10 event only showed SEBZ formation at higher pressures, the non-detection of a SEB cooling at 18 $\mu$m is unsurprising.  Observations near 20 $\mu$m should have been sufficient to resolve a cool SEBZ, and indeed Pioneer 10 measured colder SEB 20-$\mu$m temperatures in 1973 (during the faded state) than Pioneer 11 in December 1974 \citep[preceding the July 1975 revival,][]{81orton}.   However, given the low spatial and temporal resolution of these early observations we cannot confirm that the formation of an SEBZ always precedes the visible whitening.  




\subsection{Modified SEB Flow}

The first detectable sign of the onset of a fade was the cessation of the turbulent rifting to the northwest of the GRS (which normally occurs where the retrograde flow at 17$^\circ$S is deflected northwards to meet prograde flow at 7$^\circ$S).  These convective disturbances are completely absent during the faded SEB state, which may imply that heat flow from deeper levels (thought to drive convective instabilities) is no longer as efficient.  We might speculate that (i) a shift in the latitude of the GRS or (ii) a modification to the vertical extent of the vortex (and thus the extent of the blockage to the easterly and westerly flows) could modify the interaction of the GRS with the surrounding wind field at depth, preventing the discrete chaotic structures from forming.  The SEBZ formation and associated fade could then be viewed as an alternative method for transporting energy out of Jupiter's interior in the absence of discrete convective activity.  The ubiquitous subsidence over the SEB that normally keeps this region free of opacity sources \citep[see, e.g.,][]{10fletcher_grs} may then decline, permitting the development of the cool SEBZ, enhanced NH$_3$ condensation and the ultimate whitening of the belt.

The central latitude of the GRS is difficult to measure in amateur visible images due to limb-darkening effects, but limb fitting and latitude determinations between 2008 and 2010 suggested that any shifts must be smaller than 0.5$^\circ$, the standard deviation of the mean GRS planetocentric latitude (19.5$^\circ$S) as measured by the JUPOS Team\footnote{The database for Jupiter object positions project, www.jupos.org} (\textit{pers. comms.}, see acknowledgements).  Nevertheless, any shifts in the deep flows could also serve to reconfigure the retrograde SEBs jet into a meandering current, possibly giving rise to the series of five brown barges as near-stationary eddies on the northern side of the jet \citep{10rogers}.   In light of this hypothetical relationship between the dynamics of the GRS and the SEB life cycle, modelling efforts that correctly reproduce the turbulent `wake' of the GRS with a generalised moist convection scheme (particularly the addition of H$_2$O-driven convection, which has a considerable effect on the dynamics), as well as including variable blockages to the mean flow, are required to interpret the cessation of convective activity.   


Finally, we can speculate that the cessation of turbulence reflects a deeper change to jovian heat flow and meteorology on long timescales.  The existence of such deep meteorological changes \citep[known as `global upheavals,'][]{95rogers, 07rogers} has been suggested to account for the fact that SEB cycles are often associated with equally large-scale changes in the equatorial zone and in the NTB (as in 2007, though not in 2010), but no physical interpretation has yet been suggested.

\subsection{Expectations for the SEB Revival}

Based on historical records, the faded SEB could persist for 1 to 3 years \citep{95rogers, 96sanchez_jup}, and we might expect a temperature rise \citep{81orton, 92yanamandra} prior to the SEB disturbance that will ultimately restore the SEB to its usual dark brown appearance.  Such a temperature rise would raise the condensation level to higher altitudes, and lead to resublimation of the bright ices from their condensation nuclei.  If those nuclei contain the blue-absorbing chromophore responsible for the brown coloration of the belt \citep{86west}, then this mechanism would predict a slow, gradual revival.  However, visible observations show that previous SEB revivals are anything but gradual.  Instead, they are complex, violent and eruptive in nature, spreading from a single localised source called the SEB disturbance \citep[SEBD,][]{95rogers, 96sanchez_jup}.  This source is similar to convective outbreaks that occur frequently in the normal state of the SEB, but is more energetic, either because the fade process enhances the energy supply available for a subsequent outbreak, or because it provides conditions that make the outbreak self-amplifying.  From mass conservation, convective eruptions would be associated with nearby regions of local subsidence, which would cause temperatures to rise and NH$_3$ ices to sublimate, revealing the typical brown colouration of the jovian belt.  The revival then proceeds by advection of these convective plumes and subsiding regions \citep[the dark `columns' observed during previous SEB revivals,][]{95rogers} along the zonal currents within the SEB, breaking up the thickened cloud deck.  In this latter scenario, a global temperature rise across the SEB during a revival seems unlikely, and temperature enhancements are more likely to be localised in subsiding regions.

Local subsidence, atmospheric heating and `drying' (sublimation of the NH$_3$ ice coating) is likely to be responsible for the brown barges observed in the faded SEB(S) (June 2009-June 2010).  The longitudinal regularity of the subsidence within the 5 barges may be the manifestation of a zonally-propagating wave, displacing the white NH$_3$-coated aerosols to deeper, warmer pressures where the coating sublimates, revealing the darker colour. The 2007 SEBD (after the partial fade) may have erupted from a brown spot analogous to the barges observed in 2009-10 \citep{07rogers}, and the same may have occurred in 1993 \citep{96sanchez_jup} and some previous revivals \citep{95rogers}.  If so, the residual warm cyclonic circulations observed in the SEB(S) after the brown barges had faded may have an important role to play in the upcoming revival.

In summary, the VISIR observations of the 2009-10 fade are consistent with enhanced opacity and a colder SEBZ resulting from upwelling of NH$_3$-laden air.  Enhanced condensation obscures the blue-absorbing chromophore and whitens the SEB.  We predict that the next SEB disturbance and revival will not be preceded by a ubiquitous warming of the whole belt, but that convective outbreaks will have associated regions of warm, dry subsidence, re-sublimating the NH$_3$ ice and once again revealing Jupiter's typical brown SEB.  

%

%

\section{Conclusions}

Mid-infrared imaging from VLT/VISIR has been used to retrieve variations of Jupiter's atmospheric temperature and aerosol opacity during a fade of the South Equatorial Belt (SEB), the most complete whitening of this region since 1992.  By comparing the mid-IR observations with IRTF images at 4.8 $\mu$m and amateur visible images, we can draw the following conclusions:

\begin{enumerate}

\item A dramatic increase in the infrared optical thickness of SEB aerosols occurred in June-August 2009 across a range of cloud levels in the 0.8 to 3-bar range (attenuating thermal emission at 4.8 and 8.6 $\mu$m).  This preceded the increased visible reflectivity of the SEB by several months.  The visible fade of the SEB(S) started west of the GRS, and had reached the region east of the GRS by October-November 2009.  By the completion of the fade in April-July 2010, the tropospheric opacity of SEB aerosols near 800 mbar had increased by 80\% compared to their 2008 value, obscuring both the SEB(S) and the five brown barges that had been present since June 2009.  The faded SEB appeared to be remarkably quiescent (i.e., no convective plumes or spots), and only the narrow SEB(N) remained relatively cloud-free.

\item A low-temperature zone formed in the centre of the SEB at $p>300$ mbar in July-August 2009, after the cessation of turbulent convective activity northwest of the GRS in May 2009, but before the increases in visible albedo.   The SEBZ formation was only observed in the convectively-unstable region of the troposphere, but it produced a reduction in the meridional temperature gradient (and hence the thermal windshear on zonal jets) in the 100-300 mbar region.  The SEB fade had no effect on stratospheric temperatures near 5 mbar.
 
\item Condensation of NH$_3$ ice, supplied by SEBZ upwelling of air laden with condensible volatiles from the deep troposphere, is the leading candidate for the increased opacity of the 800-mbar clouds (obscuring 8.6-$\mu$m emission).  The enhanced condensation then obscured the blue-absorbing chromophore and whitened the SEB.  However, different spatial morphologies of aerosol opacity at 4.8 and 8.6 $\mu$m suggest the enhanced opacity of a second, deeper cloud deck (above the 2-3 bar level), possibly consisting of NH$_4$SH ice, although neither ice has been identified spectroscopically.  The enhanced condensation hypothesis is consistent with studies of visible reflectivity during previous SEB life cycles.

\item The cessation of turbulent convective activity northwest of the GRS preceded the formation of the SEBZ, brown barges and enhanced aerosol opacity.  The quiescent state of the `GRS wake' before the onset of the fade suggests a dynamical connection between the atmospheric flows around the GRS and the SEB life cycle.

\item Five brown barges formed in the SEB(S) in June-August 2009 as it began to fade to a pale orange.  These warm cloud-free regions of atmospheric subsidence were almost undetectable by July 2010, but their locations were still marked by warm tropospheric temperatures and faint blue patches in visible light.  Historical records suggest that these cyclonic circulations may play an important role in the next SEB revival.

\end{enumerate}

The revival of the dark coloration of Jupiter's SEB must involve sublimation of the ices from their condensation nuclei, driven by subsidence and warming of dry (volatile-depleted) air.  This dry subsidence is likely to warm the atmosphere in localised regions (dark `columns') adjacent to convective plumes (bright spots) erupting during the SEB disturbance, presumably driven by instabilities and moist convection at depth.  At the time of writing (October 2010), amateurs and professionals alike are monitoring the planet, eagerly awaiting the initial disturbance that will ultimately restore Jupiter's SEB.

Future observations of an SEB fade and revival sequence should focus on low to moderate resolution mid-infrared spectroscopy (5-25 $\mu$m), spatially resolving the SEB to solve the ambiguity between temperatures, gaseous NH$_3$ and aerosols inherent in a mid-IR filtered imaging retrieval.  Regular monitoring of the life cycles of Jupiter's axisymmetric bands should be a priority for future exploration of Jupiter, given that their different fade and revival cycles have the potential to reveal new information about the formation of belts and zones within Jupiter's dynamic troposphere.

\section*{Acknowledgments}

Fletcher was supported during this research by a Glasstone Science Fellowship at the University of Oxford.  We wish to thank the director and staff of the ESO Very Large Telescope for their assistance with the execution of these observations.  This investigation was based on observations acquired at the Paranal UT3/Melipal Observatory under Jupiter observing programmes executed between May 2008 and July 2010 (ID in Table \ref{tab:data}).  We are extremely grateful to all the amateur observers who have contributed to the ongoing coverage of Jupiter, especially those whose images are shown in the Figures: M. Salway, A. Wesley, C. Go, J.P. Prost, T. Barry and R. Chavez.  We acknowledge the JUPOS team (Hans-Joerg Mettig, Michel Jacquesson, Gianluigi Adamoli, Marco Vedovato), whose measurements of these images enabled the comparative mapping of these features.  VLT and IRTF data were reduced with the assistance of a number of JPL student interns, including E. Otto, N. Reshetnikov, A. Allahverdi, J. Greco, Z. Greene, D. Holt, S. Lai and G. Villar.  Furthermore, we thank M. Lystrup for her generous donation of time on the IRTF during the July-August observations.   Orton and Yanamandra-Fisher carried out part of this research at the Jet Propulsion Laboratory, California Institute of Technology, under a contract with NASA.  We are grateful to the staff and telescope operators at the NASA Infrared Telescope Facility operated by the University of Hawaii under Cooperative Agreement No. NCC 5-538 with the NASA Science Mission Directorate of Planetary Astronomy Program.  We are grateful to two anonymous reviewers for their constructive comments about this manuscript.


\bibliographystyle{harvard}

\bibliography{references_master}

\begin{thebibliography}{44}
\expandafter\ifx\csname natexlab\endcsname\relax\def\natexlab#1{#1}\fi
\expandafter\ifx\csname url\endcsname\relax
  \def\url#1{\texttt{#1}}\fi
\expandafter\ifx\csname urlprefix\endcsname\relax\def\urlprefix{URL }\fi

\bibitem[{{Atreya} et~al.(1999){Atreya}, {Wong}, {Owen}, {Mahaffy}, {Niemann},
  {de Pater}, {Drossart}, and {Encrenaz}}]{99atreya}
{Atreya}, S.~K., {Wong}, M.~H., {Owen}, T.~C., {Mahaffy}, P.~R., {Niemann},
  H.~B., {de Pater}, I., {Drossart}, P., {Encrenaz}, T., 1999. {A comparison of
  the atmospheres of Jupiter and Saturn: deep atmospheric composition, cloud
  structure, vertical mixing, and origin}. Plan. \& Space Sci. 47, 1243--1262.

\bibitem[{Baines et~al.(2002)Baines, Carlson, and Kamp}]{02baines}
Baines, K., Carlson, R., Kamp, L., 2002. {Fresh Ammonia Ice Clouds in Jupiter
  I. Spectroscopic Identification, Spatial Distribution, and Dynamical
  Implications}. Icarus 159~(1), 74--94.

\bibitem[{{Baines} et~al.(2007){Baines}, {Simon-Miller}, {Orton}, {Weaver},
  {Lunsford}, {Momary}, {Spencer}, {Cheng}, {Reuter}, {Jennings}, {Gladstone},
  {Moore}, {Stern}, {Young}, {Throop}, {Yanamandra-Fisher}, {Fisher}, {Hora},
  and {Ressler}}]{07baines}
{Baines}, K.~H., {Simon-Miller}, A.~A., {Orton}, G.~S., {Weaver}, H.~A.,
  {Lunsford}, A., {Momary}, T.~W., {Spencer}, J., {Cheng}, A.~F., {Reuter},
  D.~C., {Jennings}, D.~E., {Gladstone}, G.~R., {Moore}, J., {Stern}, S.~A.,
  {Young}, L.~A., {Throop}, H., {Yanamandra-Fisher}, P., {Fisher}, B.~M.,
  {Hora}, J., {Ressler}, M.~E., Oct. 2007. {Polar Lightning and Decadal-Scale
  Cloud Variability on Jupiter}. Science 318, 226--228.

\bibitem[{{Cheng} et~al.(2008){Cheng}, {Simon-Miller}, {Weaver}, {Baines},
  {Orton}, {Yanamandra-Fisher}, {Mousis}, {Pantin}, {Vanzi}, {Fletcher},
  {Spencer}, {Stern}, {Clarke}, {Mutchler}, and {Noll}}]{08cheng}
{Cheng}, A.~F., {Simon-Miller}, A.~A., {Weaver}, H.~A., {Baines}, K.~H.,
  {Orton}, G.~S., {Yanamandra-Fisher}, P.~A., {Mousis}, O., {Pantin}, E.,
  {Vanzi}, L., {Fletcher}, L.~N., {Spencer}, J.~R., {Stern}, S.~A., {Clarke},
  J.~T., {Mutchler}, M.~J., {Noll}, K.~S., Jun. 2008. {Changing Characteristics
  of Jupiter's Little Red SPOT}. Astronomical Journal 135, 2446--2452.

\bibitem[{{Flasar} et~al.(2004){Flasar}, {Kunde}, {Achterberg}, {Conrath},
  {Simon-Miller}, {Nixon}, {Gierasch}, {Romani}, {B{\'e}zard}, {Irwin},
  {Bjoraker}, {Brasunas}, {Jennings}, {Pearl}, {Smith}, {Orton}, {Spilker},
  {Carlson}, {Calcutt}, {Read}, {Taylor}, {Parrish}, {Barucci}, {Courtin},
  {Coustenis}, {Gautier}, {Lellouch}, {Marten}, {Prang{\'e}}, {Biraud},
  {Fouchet}, {Ferrari}, {Owen}, {Abbas}, {Samuelson}, {Raulin}, {Ade},
  {C{\'e}sarsky}, {Grossman}, and {Coradini}}]{04flasar_jupiter}
{Flasar}, F.~M., {Kunde}, V.~G., {Achterberg}, R.~K., {Conrath}, B.~J.,
  {Simon-Miller}, A.~A., {Nixon}, C.~A., {Gierasch}, P.~J., {Romani}, P.~N.,
  {B{\'e}zard}, B., {Irwin}, P., {Bjoraker}, G.~L., {Brasunas}, J.~C.,
  {Jennings}, D.~E., {Pearl}, J.~C., {Smith}, M.~D., {Orton}, G.~S., {Spilker},
  L.~J., {Carlson}, R., {Calcutt}, S.~B., {Read}, P.~L., {Taylor}, F.~W.,
  {Parrish}, P., {Barucci}, A., {Courtin}, R., {Coustenis}, A., {Gautier}, D.,
  {Lellouch}, E., {Marten}, A., {Prang{\'e}}, R., {Biraud}, Y., {Fouchet}, T.,
  {Ferrari}, C., {Owen}, T.~C., {Abbas}, M.~M., {Samuelson}, R.~E., {Raulin},
  F., {Ade}, P., {C{\'e}sarsky}, C.~J., {Grossman}, K.~U., {Coradini}, A., Jan.
  2004. {An intense stratospheric jet on Jupiter}. Nature 427, 132--135.

\bibitem[{{Fletcher} et~al.(2011){Fletcher}, {Orton}, {de Pater}, {Edwards},
  {Yanamandra-Fisher}, {Hammel}, and {Lisse}}]{11fletcher_trecs}
{Fletcher}, L.~N., {Orton}, G.~S., {de Pater}, I., {Edwards}, M.,
  {Yanamandra-Fisher}, P., {Hammel}, H.~B., {Lisse}, C.~M., 2011. {The
  Aftermath of the July 2009 Impact on Jupiter: Ammonia, Temperatures and
  Particulates from Gemini Thermal Infrared Spectroscopy}. Icarus 211,
  568--586.

\bibitem[{{Fletcher} et~al.(2010{\natexlab{a}}){Fletcher}, {Orton}, {de Pater},
  and {Mousis}}]{10fletcher_cxhy}
{Fletcher}, L.~N., {Orton}, G.~S., {de Pater}, I., {Mousis}, O., Dec.
  2010{\natexlab{a}}. {Jupiter's stratospheric hydrocarbons and temperatures
  after the July 2009 impact from VLT infrared spectroscopy}. Astron.
  Astrophys. 524, A46+.

\bibitem[{{Fletcher} et~al.(2010{\natexlab{b}}){Fletcher}, {Orton}, {Mousis},
  {Yanamandra-Fisher}, {Parrish}, {Irwin}, {Fisher}, {Vanzi}, {Fujiyoshi},
  {Fuse}, {Simon-Miller}, {Edkins}, {Hayward}, and {De
  Buizer}}]{10fletcher_grs}
{Fletcher}, L.~N., {Orton}, G.~S., {Mousis}, O., {Yanamandra-Fisher}, P.,
  {Parrish}, P.~D., {Irwin}, P.~G.~J., {Fisher}, B.~M., {Vanzi}, L.,
  {Fujiyoshi}, T., {Fuse}, T., {Simon-Miller}, A.~A., {Edkins}, E., {Hayward},
  T.~L., {De Buizer}, J., Jul. 2010{\natexlab{b}}. {Thermal structure and
  composition of Jupiter's Great Red Spot from high-resolution thermal
  imaging}. Icarus 208, 306--328.

\bibitem[{{Fletcher} et~al.(2009{\natexlab{a}}){Fletcher}, {Orton}, {Teanby},
  and {Irwin}}]{09fletcher_ph3}
{Fletcher}, L.~N., {Orton}, G.~S., {Teanby}, N.~A., {Irwin}, P.~G.~J., Aug.
  2009{\natexlab{a}}. {Phosphine on Jupiter and Saturn from Cassini/CIRS}.
  Icarus 202, 543--564.

\bibitem[{{Fletcher} et~al.(2009{\natexlab{b}}){Fletcher}, {Orton},
  {Yanamandra-Fisher}, {Fisher}, {Parrish}, and {Irwin}}]{09fletcher_imaging}
{Fletcher}, L.~N., {Orton}, G.~S., {Yanamandra-Fisher}, P., {Fisher}, B.~M.,
  {Parrish}, P.~D., {Irwin}, P.~G.~J., Mar. 2009{\natexlab{b}}. {Retrievals of
  atmospheric variables on the gas giants from ground-based mid-infrared
  imaging}. Icarus 200, 154--175.

\bibitem[{{Friedson}(1999)}]{99friedson}
{Friedson}, A.~J., Jan. 1999. {New Observations and Modelling of a QBO-Like
  Oscillation in Jupiter's Stratosphere}. Icarus 137, 34--55.

\bibitem[{{Ingersoll} et~al.(2004){Ingersoll}, {Dowling}, {Gierasch}, {Orton},
  {Read}, {Sanchez-Lavega}, {Showman}, {Simon-Miller}, and
  {Vasavada}}]{04ingersoll}
{Ingersoll}, A.~P., {Dowling}, T.~E., {Gierasch}, P.~J., {Orton}, G.~S.,
  {Read}, P.~L., {Sanchez-Lavega}, A., {Showman}, A.~P., {Simon-Miller}, A.~A.,
  {Vasavada}, A.~R., 2004. {Dynamics of Jupiter's atmosphere}. Jupiter.~The
  Planet, Satellites and Magnetosphere, pp. 105--128.

\bibitem[{Irwin et~al.(2008)Irwin, Teanby, de~Kok, Fletcher, Howett, Tsang,
  Wilson, Calcutt, Nixon, and Parrish}]{08irwin}
Irwin, P., Teanby, N., de~Kok, R., Fletcher, L., Howett, C., Tsang, C., Wilson,
  C., Calcutt, S., Nixon, C., Parrish, P., 2008. {The NEMESIS planetary
  atmosphere radiative transfer and retrieval tool}. Journal of Quantitative
  Spectroscopy and Radiative Transfer 109~(6), 1136--1150.

\bibitem[{{Kuehn} and {Beebe}(1993)}]{93kuehn}
{Kuehn}, D.~M., {Beebe}, R.~F., Feb. 1993. {A study of the time variability of
  Jupiter's atmospheric structure}. Icarus 101, 282--292.

\bibitem[{{Lagage} et~al.(2004){Lagage}, {Pel}, {Authier}, {Belorgey},
  {Claret}, {Doucet}, {Dubreuil}, {Durand}, {Elswijk}, {Girardot}, {K{\"a}ufl},
  {Kroes}, {Lortholary}, {Lussignol}, {Marchesi}, {Pantin}, {Peletier},
  {Pirard}, {Pragt}, {Rio}, {Schoenmaker}, {Siebenmorgen}, {Silber}, {Smette},
  {Sterzik}, and {Veyssiere}}]{04lagage}
{Lagage}, P.~O., {Pel}, J.~W., {Authier}, M., {Belorgey}, J., {Claret}, A.,
  {Doucet}, C., {Dubreuil}, D., {Durand}, G., {Elswijk}, E., {Girardot}, P.,
  {K{\"a}ufl}, H.~U., {Kroes}, G., {Lortholary}, M., {Lussignol}, Y.,
  {Marchesi}, M., {Pantin}, E., {Peletier}, R., {Pirard}, J., {Pragt}, J.,
  {Rio}, Y., {Schoenmaker}, T., {Siebenmorgen}, R., {Silber}, A., {Smette}, A.,
  {Sterzik}, M., {Veyssiere}, C., Sep. 2004. {Successful Commissioning of
  VISIR: The Mid-Infrared VLT Instrument}. The Messenger 117, 12--16.

\bibitem[{{Leovy} et~al.(1991){Leovy}, {Friedson}, and {Orton}}]{91leovy}
{Leovy}, C.~B., {Friedson}, A.~J., {Orton}, G.~S., Dec. 1991. {The
  quasiquadrennial oscillation of Jupiter's equatorial stratosphere}. Nature
  354, 380--382.

\bibitem[{Matcheva et~al.(2005)Matcheva, Conrath, Gierasch, and
  Flasar}]{05matcheva}
Matcheva, K., Conrath, B., Gierasch, P., Flasar, F., 2005. {The cloud structure
  of the jovian atmosphere as seen by the Cassini/CIRS experiment}. Icarus
  179~(2), 432--448.

\bibitem[{{Moreno} et~al.(1997){Moreno}, {Molina}, and {Ortiz}}]{97moreno}
{Moreno}, F., {Molina}, A., {Ortiz}, J.~L., Nov. 1997. {The 1993 south
  equatorial belt revival and other features in the Jovian atmosphere: an
  observational perspective}. Astron. Astrophys. 327, 1253--1261.

\bibitem[{{Nixon} et~al.(2007){Nixon}, {Achterberg}, {Conrath}, {Irwin},
  {Teanby}, {Fouchet}, {Parrish}, {Romani}, {Abbas}, {Leclair}, {Strobel},
  {Simon-Miller}, {Jennings}, {Flasar}, and {Kunde}}]{07nixon}
{Nixon}, C.~A., {Achterberg}, R.~K., {Conrath}, B.~J., {Irwin}, P.~G.~J.,
  {Teanby}, N.~A., {Fouchet}, T., {Parrish}, P.~D., {Romani}, P.~N., {Abbas},
  M., {Leclair}, A., {Strobel}, D., {Simon-Miller}, A.~A., {Jennings}, D.~J.,
  {Flasar}, F.~M., {Kunde}, V.~G., May 2007. {Meridional variations of
  C$_{2}$H$_{2}$ and C$_{2}$H$_{6}$ in Jupiter's atmosphere from Cassini CIRS
  infrared spectra}. Icarus 188, 47--71.

\bibitem[{Orton et~al.(2011)Orton, Fletcher, Lisse, Chodas, Cheng,
  Yanamandra-Fisher, Baines, Fisher, Wesley, Perez-Hoyos, de~Pater, Hammel,
  Edwards, Mousis, Marchis, Golisch, Sanchez-Lavega, Simon-Millerl, Hueso,
  Momary, Greene, Reshetnikov, Otto, Villar, Lai, and Wong}]{11orton}
Orton, G.~S., Fletcher, L.~N., Lisse, C.~M., Chodas, P.~W., Cheng, A.,
  Yanamandra-Fisher, P.~A., Baines, K.~H., Fisher, B.~M., Wesley, A.,
  Perez-Hoyos, S., de~Pater, I., Hammel, H.~B., Edwards, M.~L., Mousis, O.,
  Marchis, F., Golisch, W., Sanchez-Lavega, A., Simon-Millerl, A.~A., Hueso,
  R., Momary, T.~W., Greene, Z., Reshetnikov, N., Otto, E., Villar, G., Lai,
  S., Wong, M., 2011. {The Atmospheric Influence, Size and Asteroidal Nature of
  the July 2009 Jupiter Impactor}. Icarus 211, 587--602.

\bibitem[{{Orton} et~al.(1994){Orton}, {Friedson}, {Yanamandra-Fisher},
  {Caldwell}, {Hammel}, {Baines}, {Bergstralh}, {Martin}, {West}, {Veeder},
  {Lynch}, {Russell}, {Malcom}, {Golisch}, {Griep}, {Kaminski}, {Tokunaga},
  {Herbst}, and {Shure}}]{94orton}
{Orton}, G.~S., {Friedson}, A.~J., {Yanamandra-Fisher}, P.~A., {Caldwell}, J.,
  {Hammel}, H.~B., {Baines}, K.~H., {Bergstralh}, J.~T., {Martin}, T.~Z.,
  {West}, R.~A., {Veeder}, Jr., G.~J., {Lynch}, D.~K., {Russell}, R., {Malcom},
  M.~E., {Golisch}, W.~F., {Griep}, D.~M., {Kaminski}, C.~D., {Tokunaga},
  A.~T., {Herbst}, T., {Shure}, M., Jul. 1994. {Spatial Organization and Time
  Dependence of Jupiter's Tropospheric Temperatures, 1980-1993}. Science 265,
  625--631.

\bibitem[{{Orton} et~al.(1981){Orton}, {Ingersoll}, {Terrile}, and
  {Walton}}]{81orton}
{Orton}, G.~S., {Ingersoll}, A.~P., {Terrile}, R.~J., {Walton}, S.~R., Aug.
  1981. {Images of Jupiter from the Pioneer 10 and Pioneer 11 Infrared
  Radiometers - A comparison with visible and 5-micron images}. Icarus 47,
  145--158.

\bibitem[{{Peek}(1958)}]{58peek}
{Peek}, B.~M., 1958. {The Planet Jupiter}. {Faber and Faber, London}.

\bibitem[{{Reuter} et~al.(2007){Reuter}, {Simon-Miller}, {Lunsford}, {Baines},
  {Cheng}, {Jennings}, {Olkin}, {Spencer}, {Stern}, {Weaver}, and
  {Young}}]{07reuter}
{Reuter}, D.~C., {Simon-Miller}, A.~A., {Lunsford}, A., {Baines}, K.~H.,
  {Cheng}, A.~F., {Jennings}, D.~E., {Olkin}, C.~B., {Spencer}, J.~R., {Stern},
  S.~A., {Weaver}, H.~A., {Young}, L.~A., Oct. 2007. {Jupiter Cloud
  Composition, Stratification, Convection, and Wave Motion: A View from New
  Horizons}. Science 318, 223--225.

\bibitem[{Rogers(1995)}]{95rogers}
Rogers, J., 1995. {The Giant Planet Jupiter}. Cambridge University Press.

\bibitem[{{Rogers}(2007{\natexlab{a}})}]{07rogers}
{Rogers}, J.~H., Jun. 2007{\natexlab{a}}. {Jupiter embarks on a 'global
  upheaval'}. Journal of the British Astronomical Association 117, 113--115.

\bibitem[{{Rogers}(2007{\natexlab{b}})}]{07rogers_climax}
{Rogers}, J.~H., Oct. 2007{\natexlab{b}}. {The climax of Jupiter's global
  upheaval}. Journal of the British Astronomical Association 117, 226--230.

\bibitem[{{Rogers}(2010{\natexlab{a}})}]{10rogers}
{Rogers}, J.~H., July 2010{\natexlab{a}}. {Jupiter in 2009: Interim Report,
  with new insights into the NTZ disturbance, NEB expansion, and SEB fading}.
\newline\urlprefix\url{http://www.britastro.org/jupiter/2009report07.htm}

\bibitem[{{Rogers}(2010{\natexlab{b}})}]{10rogers_BA}
{Rogers}, J.~H., June 2010{\natexlab{b}}. {Jupiter in 2010: BAA Jupiter Section
  Bulletin}.
\newline\urlprefix\url{http://www.britastro.org/jupiter/2010report04.htm}

\bibitem[{{Rogers} et~al.(2011){Rogers}, {Adamoli}, and
  {Mettig}}]{11rogers_LRS}
{Rogers}, J.~H., {Adamoli}, G., {Mettig}, H., 2011. Jupiter's high-latitude
  storms: A little red spot tracked through a jovian year. Journal of the
  British Astronomical Association 121~(1), 19--29.

\bibitem[{{Rogers} et~al.(2010){Rogers}, {Mettig}, {Adamoli}, {Jacquesson}, and
  {Vedovato}}]{10rogers_spots}
{Rogers}, J.~H., {Mettig}, H., {Adamoli}, G., {Jacquesson}, M., {Vedovato}, M.,
  October 2010. {Jupiter in 2010: Interim report: Southern hemisphere }.
\newline\urlprefix\url{http://www.britastro.org/jupiter/2010report08.htm}

\bibitem[{{Rogers} and {Young}(1977)}]{77rogers}
{Rogers}, J.~H., {Young}, P.~J., Apr. 1977. {Earth-based and Pioneer
  observations of Jupiter}. Journal of the British Astronomical Association 87,
  240--251.

\bibitem[{{Rogers} et~al.(2004){Rogers}, {Akutsu}, and {Orton}}]{04rogers}
{Rogers}, O.~H., {Akutsu}, T., {Orton}, G.~S., Dec. 2004. {Jupiter in
  2000/2001. Part II: Infrared and ultraviolet wavelengths}. Journal of the
  British Astronomical Association 114, 313.

\bibitem[{{Roos-Serote} et~al.(1998){Roos-Serote}, {Drossart}, {Encrenaz},
  {Lellouch}, {Carlson}, {Baines}, {Kamp}, {Mehlman}, {Orton}, {Calcutt},
  {Irwin}, {Taylor}, and {Weir}}]{98roos-serote}
{Roos-Serote}, M., {Drossart}, P., {Encrenaz}, T., {Lellouch}, E., {Carlson},
  R.~W., {Baines}, K.~H., {Kamp}, L., {Mehlman}, R., {Orton}, G.~S., {Calcutt},
  S., {Irwin}, P., {Taylor}, F., {Weir}, A., Sep. 1998. {Analysis of Jupiter
  North Equatorial Belt hot spots in the 4-5-{$\mu$}m range from
  Galileo/near-infrared mapping spectrometer observations: Measurements of
  cloud opacity, water, and ammonia}. Journal of Geophysical Research 103,
  23023--23042.

\bibitem[{{Russell} and {Dougherty}(2010)}]{10russell}
{Russell}, C.~T., {Dougherty}, M.~K., May 2010. {Magnetic Fields of the Outer
  Planets}. Space Science Reviews 152, 251--269.

\bibitem[{{Sanchez-Lavega} and {Gomez}(1996)}]{96sanchez_jup}
{Sanchez-Lavega}, A., {Gomez}, J.~M., May 1996. {The South Equatorial Belt of
  Jupiter, I: Its Life Cycle}. Icarus 121, 1--17.

\bibitem[{{Sanchez-Lavega} et~al.(1996){Sanchez-Lavega}, {Gomez}, {Lecacheux},
  {Colas}, {Miyazaki}, {Parker}, and {Guarro}}]{96sanchez_SEB}
{Sanchez-Lavega}, A., {Gomez}, J.~M., {Lecacheux}, J., {Colas}, F., {Miyazaki},
  I., {Parker}, D., {Guarro}, J., May 1996. {The South Equatorial Belt of
  Jupiter, II: The Onset and Development of the 1993 Disturbance}. Icarus 121,
  18--29.

\bibitem[{{S{\'a}nchez-Lavega} et~al.(2008){S{\'a}nchez-Lavega}, {Orton},
  {Hueso}, {Garc{\'{\i}}a-Melendo}, {P{\'e}rez-Hoyos}, {Simon-Miller}, {Rojas},
  {G{\'o}mez}, {Yanamandra-Fisher}, {Fletcher}, {Joels}, {Kemerer}, {Hora},
  {Karkoschka}, {de Pater}, {Wong}, {Marcus}, {Pinilla-Alonso}, {Carvalho},
  {Go}, {Parker}, {Salway}, {Valimberti}, {Wesley}, and {Pujic}}]{08sanchez}
{S{\'a}nchez-Lavega}, A., {Orton}, G.~S., {Hueso}, R., {Garc{\'{\i}}a-Melendo},
  E., {P{\'e}rez-Hoyos}, S., {Simon-Miller}, A., {Rojas}, J.~F., {G{\'o}mez},
  J.~M., {Yanamandra-Fisher}, P., {Fletcher}, L., {Joels}, J., {Kemerer}, J.,
  {Hora}, J., {Karkoschka}, E., {de Pater}, I., {Wong}, M.~H., {Marcus}, P.~S.,
  {Pinilla-Alonso}, N., {Carvalho}, F., {Go}, C., {Parker}, D., {Salway}, M.,
  {Valimberti}, M., {Wesley}, A., {Pujic}, Z., Jan. 2008. {Depth of a strong
  jovian jet from a planetary-scale disturbance driven by storms}. Nature 451,
  437--440.

\bibitem[{{Satoh} and {Kawabata}(1992)}]{92satoh}
{Satoh}, T., {Kawabata}, K., Jan. 1992. {Methane band photometry of the faded
  South Equatorial Belt of Jupiter}. ApJ 384, 298--304.

\bibitem[{{Satoh} and {Kawabata}(1994)}]{94satoh}
{Satoh}, T., {Kawabata}, K., Apr. 1994. {A change of upper cloud structure in
  Jupiter's South Equatorial Belt during the 1989-1990 event}. Journal of
  Geophysical Research 99, 8425--8440.

\bibitem[{{Weidenschilling} and {Lewis}(1973)}]{73weidenschilling}
{Weidenschilling}, S.~J., {Lewis}, J.~S., 1973. {Atmospheric and cloud
  structures of the jovian planets}. Icarus 20, 465--476.

\bibitem[{{West} et~al.(1986){West}, {Strobel}, and {Tomasko}}]{86west}
{West}, R.~A., {Strobel}, D.~F., {Tomasko}, M.~G., 1986. {Clouds, aerosols, and
  photochemistry in the Jovian atmosphere}. Icarus 65, 161--217.

\bibitem[{{Wong} et~al.(2004){Wong}, {Bjoraker}, {Smith}, {Flasar}, and
  {Nixon}}]{04wong}
{Wong}, M.~H., {Bjoraker}, G.~L., {Smith}, M.~D., {Flasar}, F.~M., {Nixon},
  C.~A., 2004. {Identification of the 10-{$\mu$}m ammonia ice feature on
  Jupiter}. Plan. \& Space Sci. 52, 385--395.

\bibitem[{{Yanamandra-Fisher} et~al.(1992){Yanamandra-Fisher}, {Orton}, and
  {Friedson}}]{92yanamandra}
{Yanamandra-Fisher}, P., {Orton}, G., {Friedson}, J., Jun. 1992. {Time
  Dependence of Jupiter's Tropospheric Temperatures and Cloud Properties: The
  1989 SEB Disturbance}. In: Bulletin of the American Astronomical Society.
  Vol.~24 of Bulletin of the American Astronomical Society. p. 1039.

\end{thebibliography}









\renewcommand{\baselinestretch}{1}

\clearpage
\appendix

\section{Phenomena at Other Latitudes}
\label{phenom}

\begin{figure*}[tbp]
\centering
\epsfig{file=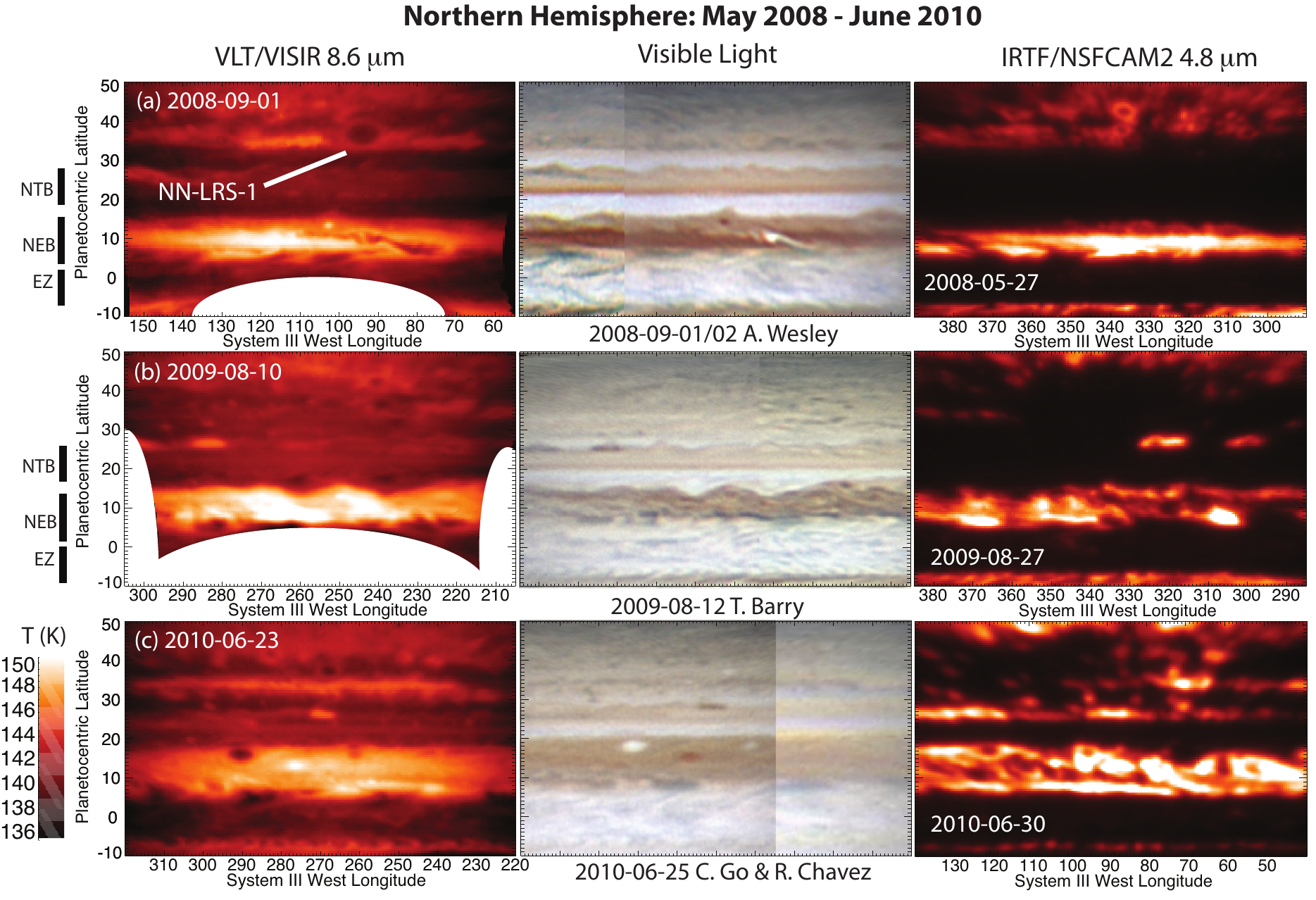,width=\textwidth,angle=0}
\caption{Phenomena in the northern hemisphere between May 2008 and June 2010, observed by VLT/VISIR at 8.6 $\mu$m (left, sensitive to aerosol opacity at $p<800$ mbar); visible light from amateur observers (centre); and IRTF/NSFCAM2 observations at 4.8 $\mu$m (right, sensitive to aerosols above the 2-3 bar level). Visible images taken as close as possible to the VISIR observations have been provided by A. Wesley, T. Barry, C. Go and R. Chavez.  Suppressed emission at both 4.8 and 8.6 $\mu$m is caused by excess aerosol opacity.   NSFCAM2 images were taken on different nights to the VISIR images, and do not show the same longitude range.  The three epochs show the northward expansion of the NEB and the cloudy conditions of the northern Little Red Spot (NN-LRS-1) in 2008 (see \ref{phenom}).  Large white arcs in the 8.6-$\mu$m images are due to the removal of negative-beam artefacts caused by the small 20" chopping amplitude and the reprojection of the edges of the VISIR array.}
\label{northmap}
\end{figure*}

The SEB life cycle is one of numerous dynamical phenomena evident in the 2008-2010 datasets, described briefly below.  These events are not necessarily connected to the SEB fade, but appear to be part of longer-term cycles or secular changes \citep[e.g.,][]{95rogers}.  We have chosen to include them for completeness.

\begin{itemize}
\item \textit{Oval BA: } Oval BA is Jupiter's second largest anticyclone, which formed from successive mergers and reddened in 2005 \citep[e.g.,][]{08cheng}.  The regular biennial passage of Oval BA past the Great Red Spot is evident in Figs. \ref{TBmaps1}-\ref{TBmaps2}.  The cyclonic activity to the northwest of Oval BA, analagous to the GRS wake, is considerably more elongated and brighter at 8.6 $\mu$m in 2010 than it was in 2008.  This was the result of convergence with a large cyclonic cell, which initiated a merger of Oval BA with a small anticyclonic white oval and concurrent convective outbreak event in the cyclonic cell in June 2010 \citep[described in detail by][]{10rogers_BA}.   This manifests itself as clear cloud-free conditions in the cyclonic region to the northwest of the oval in Fig. \ref{aermaps}c, which also shows indications that Oval BA's aerosols were less optically thick in 2010 than in 2008, although the oval remained red in both years.  Furthermore, the warm southern edge of Oval BA (Fig. \ref{tempmaps}) was more diffuse and warmer in 2008, whereas it appeared offset to the west and cooler by 2-3 K in 2010.  It is unclear whether these changes are the result of a weakening of Oval BA between 2008 and 2010, or simply due to different environmental conditions in the STB surrounding the vortex.


\item \textit{Northern Equatorial Belt: }  The visibly-dark brown NEB is the brightest band on the planet at all thermal wavelengths (Fig. \ref{northmap}, \ref{cmerid_TB}).  Despite being a symmetric counterpart of the SEB, the NEB does not display the same classical fade and revival cycle.  However, it does undergo occasional northward broadenings that may be comparable events \citep{94orton, 95rogers}, and have occurred every 3-5 years since 1988.  The latest such event occurred between in 2009:  the NEB expanded northwards from 15\degree N to 18\degree N, at visible and thermal wavelengths, causing a narrowing of the adjacent cool North Tropical Zone (NTrZ).  This expansion corresponds to a clearing of aerosol opacity between 15-18\degree N between 2008 and 2010 (Fig. \ref{5um_merid}, Fig. \ref{gas_compare}), as well as a warming of 3-4 K in the upper troposphere for the NEBn (Fig. \ref{Tcompare}).  \citet{94orton} also noted warming associated with an earlier NEB expansion.  The resulting decrease in $dT/dy$ between 15 and 20\degree N means that the prograde jet at 21\degree N is subject to smaller vertical windshear and can persist to higher altitudes in the stratosphere.  Finally, the NEB demonstrates dramatic longitudinal wave activity with a 25-30\degree wavelength in 2009 (Fig. \ref{northmap}(b)), which was not apparent in the SEB at any time during 2008-2010. The thermal waves may be associated with an alternating array of anticyclonic and cyclonic ovals observed in visible light, some of which were newly formed as part of the expansion event \citep{10rogers, 04rogers}.  


\item \textit{Northern Vortices: }  Large red anticyclones also occur in the northern hemisphere at 38$^\circ$N in the North North Temperate Zone (NNTZ).  The largest of these, named NN-LRS-1, is frequently reddish; almost as large as Oval BA; and has existed at least since 1993 \citep{11rogers_LRS}.  It is prominent as a dark oval at thermal wavelengths in Fig. \ref{northmap}a, and appears larger at 8.6 $\mu$m than it does at 10.8 $\mu$m, suggesting that the region of elevated aerosol opacity extends over a wider area than the cold temperature conditions at the centre of the storms.     This is similar to the observed properties of Oval BA itself in Figs. \ref{TBmaps1}-\ref{south10.8}.

\item \textit{North Temperate Belt: }  Besides the SEB, the warm NTB (21-28\degree N) is one of the most active regions on the planet, also demonstrating sporadic fade and revival cycles. The last such revival of the faded NTB occurred in 2007, with the appearance of two high-altitude plumes with wakes of turbulent dark brown material rapidly encircling the planet \citep{08sanchez, 07rogers}.  The revival resulted in a broad red NTB with a component of high-altitude tropospheric haze (200-400 mbar) detectable in methane-band imaging at 2.3 $\mu$m \citep{08sanchez}.  Accordingly, the NTB has remained dark at 8.6 $\mu$m between 2008 and 2010 (Fig. \ref{northmap}), indicating the continued presence of high aerosol opacity similar to the EZ and the faded SEB (Fig. \ref{gas_compare}) and making it almost as dark as the cold NTrZ (18-21$^\circ$N).   The exceptions to this rule are discrete cloud-free barges of high 8.6-$\mu$m flux, which are the darkest parts of the revived NTBn in visible light and probably coincide with features of high 4.8-$\mu$m flux (although IRTF/NSFCAM2 and VLT/VISIR did not observe simultaneously).
\end{itemize}

\end{document}